\documentclass[twocolumn, UTF8]{aastex62}
\usepackage{amsmath}
\usepackage{diagbox}
\usepackage{appendix}
\usepackage{multirow}
\usepackage{hyperref}
\usepackage{acronym}   
\usepackage{subfigure}
\usepackage{graphicx}
\hypersetup{
    colorlinks=true,
    linkcolor=blue,
    filecolor=magenta,      
    urlcolor=cyan,
    citecolor=blue,
}
\begin{document}
\title{Exploring the sky localization and early warning capabilities of third generation gravitational wave detectors in three-detector network configurations.}
\author{YUFENG LI}
\affil{Key Laboratory for Computational Astrophysics, 
National Astronomical Observatories, Chinese Academy of Sciences, 
Beijing 100101, China}
\affil{School of Astronomy and Space Science, University of Chinese Academy of Sciences, Beijing 100049, China}
\affil{SUPA, School of Physics and Astronomy, University of Glasgow, Glasgow G12 8QQ, United Kingdom}

\author{Ik Siong Heng}
\affil{SUPA, School of Physics and Astronomy, University of Glasgow, Glasgow G12 8QQ, United Kingdom}
\correspondingauthor{Ik Siong Heng}
\email{ik.heng@glasgow.ac.uk}
\author{Man Leong Chan}
\affil{Department of Applied Physics, Fukuoka University, Nanakuma 8-19-1, Fukuoka 814-0180, Japan}

\author{Chris Messenger}
\affil{SUPA, School of Physics and Astronomy, University of Glasgow, Glasgow G12 8QQ, United Kingdom}
\author{XILONG FAN}
\affil{School of Physics and Technology, Wuhan University, Wuhan 430072, China}
\begin{abstract}
This work characterises the sky localization and early warning performance of different networks of third generation gravitational wave detectors, consisting of different combinations of detectors with either the Einstein Telescope or Cosmic Explorer configuration in sites in North America, Europe and Australia.
Using a Fisher matrix method which includes the effect of earth rotation, we estimate the sky localization uncertainty for $1.4\text{M}\odot$-$1.4\text{M}\odot$
binary neutron star mergers at distances $40\text{Mpc}$, $200\text{Mpc}$, $400\text{Mpc}$, $800\text{Mpc}$, $1600\text{Mpc}$, to characterize each network's performance for binary neutron star observations at a given distance.
We also characterize each network's sky localization capability for an assumed astrophysical population up to redshift $\leq 2$.  
Furthermore, we also study the capabilities for the different networks to localize a binary neutron star merger prior to merger (early warning) and characterise the network performance for sky localization uncertainty between 1 and 30 square degrees.
We find that, for example, for binary neutron star mergers at $200\text{Mpc}$ and a network consisting of the Einstein Telescope, 
Cosmic Explorer and an extra Einstein Telescope-like detector in Australia (2ET1CE), the upper limit of the size of the 90\% credible region for the best localized 90\% signals is $0.25\text{deg}^2$.  For the simulated astrophysical distribution,  this upper limit is $91.79\text{deg}^2$. If the Einstein Telescope-like detector in Australia is replaced with a Cosmic Explorer-like detector (1ET2CE), for signals at $200\text{Mpc}$, the size of the 90\% credible region for the best localized 90\% signals is $0.18\text{deg}^2$, while the corresponding value for the best localized 90\% sources following the astrophysical distribution is $56.77\text{deg}^2$. 
We note that the 1ET2CE network can detect 7.2$\%$ more of the simulated astrophysical population than the 2ET1CE network.
In terms of early warning performance (e.g. 200 Mpc), we find that a network of 2ET1CE and 1ET2CE networks can both provide early warnings of the order of 1 hour prior to merger with sky localization uncertainties of 30 square degrees or less. In some cases, the 2ET1CE network is capable of estimating the sky location with an uncertainty of 5 square degrees or less on timescales of about 1 hour prior to merger. 
Our study concludes that the 1ET2CE network is a good compromise between binary neutron stars detection rate, sky localization and early warning capabilities. 
\end{abstract}
\acrodef{EM}[EM]{electromagnetic}
\acrodef{ET}[ET]{Einstein Telescope}
\acrodef{CE}[CE]{Cosmic Explorer}
\acrodef{GW}[GW]{gravitational wave}
\acrodef{BNS}[BNS]{binary neutron star}
\acrodef{BBH}[BBH]{binary black hole}
\acrodef{NSBH}[NSBH]{neutron star black hole}
\acrodef{CBC}[CBC]{compact binary coalescence}
\acrodef{SGRB}[SGRB]{short-duration gamma-ray burst}
\acrodef{SNR}[SNR]{signal-to-noise-ratio}
\acrodef{DTD}[DTD]{delay time distribution}
\section{Introduction}
The first directly detected \ac{GW} event GW150914 by Advanced LIGO in September, 2015 opened a new window on exploring the Universe~\citep{2016PRL...116...061102L}.  
During the first two observation runs (O1 and O2) of Advanced LIGO, and Advanced Virgo (joined in O2)~\citep{2015CQGra..32b4001A}, 
a total of ten binary black holes (BBHs) mergers and one binary neutron star (BNS) merger were identified~\citep{2019PhRvX...9c1040A}. 
The follow-up observations of the \ac{EM} counterparts of the BNS event have initiated the GW and EM multi-messenger astronomy era~\citep{2017PhRvL.119p1101A}. 
In 2019, the Advanced LIGO and Advanced Virgo started the third observation run (O3) with improved sensitivities. O3 has finished and 39 GW candidates from the first half of O3 have been released in 2020 as the second Gravitational-Wave Transient Catalog~(GWTC-2)~\citep{2020arXiv201014527A}. More recently, a deeper list of candidate events over the same period of GWTC-2 was reported recently in \citet{2021arXiv210801045T} as GWTC-2.1, which revealed 44 high-significance candidates. It is expected that a large number of \ac{GW} events will come into our view with the improvement in sensitivity of upcoming the fourth observation run and more distant third generation detector era. 

\indent  In the frequency band relevant to ground based \ac{GW} detectors, the most common sources of \ac{GW} events are expected to be \acp{CBC}.   \\
In addition to the exploration of the source properties for a single \ac{CBC} system, GW detections could also shed light on the merger rate of \ac{CBC} system as a function of distance from us, thus could test theoretical models for \ac{CBC} merger rate and further give clues to studies of formation of single \ac{CBC} system.  For example, the BNS merger rate is constrained to be $1540^{+3200}_{-1220} \mathrm{Gpc}^{-3} \mathrm{yr}^{-1}$ by the detection of GW170817~\citep{2017PhRvL.119p1101A}; and the latest BNS merger rate is found to be 320$^{+490}_{-240}\mathrm{Gpc}^{-3} \mathrm{yr}^{-1}$ by the second LIGO-Virgo Gravitational-Wave Transient Catalog~\citep{2021ApJ...913L...7A}.
Beyond that, a fundamental method to estimate merger rate distribution, \ac{DTD}, attracts some attempts to evaluate its validity by current second-generation \ac{GW} detectors and future third generation \ac{GW} detectors~\citep{2019ApJ...878L..12S, 2019ApJ...878L..13S, 2019ApJ...878L..14S}.  As explored in a series researches not limited to ~\citet{2008ApJ...683L.127H,  1992ApJ...389L..45P,2019ApJ...878L..12S, 2019ApJ...878L..13S, 2019ApJ...878L..14S, 2007ApJ...665.1220Z}, the merger rate distribution is a convolution of the cosmic star formation rate and the distribution of delay time (\ac{DTD}) between the system form and merger.  Specifically, the \ac{DTD} is usually parametrized as a power law distribution with a slope $\Gamma$ above minimum merger timescale $t_\text{{min}}$. In later analysis, we generated \ac{GW} sources following \ac{DTD} distribution to simulate a relative realistic distribution in the Universe. \\
\indent Systems of \acp{CBC} which include at least one neutron star, such as the mergers of \ac{BNS} or \ac{NSBH}, are likely to generate observable \ac{EM} emissions. 
These emissions include \acp{SGRB}, powered by accretion onto the central compact object~\citep{2012ApJ...746...48M}, and an
isotropic thermal emission powered by the radioactive decay of heavy elements in the merger ejecta, known as kilonova \citep{2010MNRAS.406.2650M}, as well as optical and radio afterglows~\citep{2019ApJ...886L..17H}. 
Successful \ac{EM} follow-up observations of a \ac{GW} event can identify the host galaxy, and provide information 
on the progenitor local environment and the hydrodynamics of the merger~\citep{2017ApJ...850L..40A, 2016JPhCS.718b2004B, 2010PhRvD..82j2002N}. 
Coupled with the estimates of 
a model-independent luminosity distance derived from \acp{GW},
it is even possible to infer the values of the Hubble constant~\citep{2021ApJ...909..218A,2019NatAs...3..940H, 2018arXiv181204051M}.
In terms of \ac{EM} follow-up observations, the most needed information from GWs is the estimate of the sky location of the source, the size of the error region, and the time at merger.  To increase the likelihood of successful \ac{EM} follow-up observations, therefore, improvement in localization  of \acp{GW} is desired. 
 On the other hand, if the in-band duration of a \ac{GW} signal is long enough that its presence can be identified and information on its sky location obtained 
 well before the merger, astronomers will be allowed more time to prepare, further increasing the likelihood of successful \ac{EM} follow-up observations. 
 This is known as early warning~\citep{2012ApJ...748..136C}.  For \acp{BNS}, the in-band duration of a signal mainly depends on the low cut-off frequency of \ac{GW} detectors and the masses of the systems. For example, the in-band duration of the insprial signal of a $1.4M_\odot-1.4M_\odot$ \ac{BNS} meger in a \ac{GW} detector with a low cut-off frequency of $10$ Hz
 is $\mathcal{O}(10^3)$s.  \\
 \indent Third generation GW detectors such as the \ac{ET}~\citep{2014ASSL..404..333P} and 
\ac{CE}~\citep{2017CQGra..34d4001A} are expected to be built in the 2030s. 
With sentivities better than that of the second generation detectors,  observing \ac{GW}s from \ac{BBH} and \ac{BNS} sources located at
distances far greater than the horizon of second generation detectors will be possible, thus could contribute to explorations of compact object evolution so as to further study the evolution of the Universe. Localization of \ac{CBC} sources detectable by the second generation detectors and \ac{CBC} sources at greater distances could also benefit from improved sensitivity of the third generation detectors due to higher SNR.  And the improved sensitivity of third generation \ac{GW} detectors at low frequency band, for example, 1~Hz-10~Hz, allows for a longer in-band duration (e.g. low cut-off frequency 1~Hz, in-band duration will be~5 days) for \ac{CBC} signals as the time to merger is inversely related to the instantaneous \ac{GW} frequency which we will introduce later, thus jointly lead to a promising EM follow-up observations as stated in previous researches such as \citet{2018arXiv180810057A, 2018arXiv181207307A, 2018PhRvD..97l3014C, 2018PhRvD..97j4064M, 2018PhRvD..97f4031Z}. It should be noted that we considered the earth's rotation due to the long duration of simulated signals.\\
 \indent In this paper, using a Fisher matrix approach, we estimate the localization uncertainty and early warning prospect for networks of third generation ground based \ac{GW}
detectors for \acp{BNS}. 
We simulate \ac{BNS} mergers at fixed distances as well as a population following the \ac{DTD} of a power law distribution with a slope $\Gamma = -1.5$ 
and a minimum merger time $t_\mathrm{min} = 1$ Gyr. 

Previous studies have shown the benefits of building a detector in Australia because of the long baseline that such a detector will form with other detectors~\citep{2010CQGra..27h4005B,2018MNRAS.474.4385H, 2011ApJ...739...99N}. This paper aims to characterise the performance of a network of 3 third generation detectors which consists of different permutations of detectors in CE and ET configurations.  Thus, in addition to the proposed third generation \ac{GW} detectors, the \ac{ET} and \ac{CE},  we assume an ET-like detector or a CE-like detector in Australia.\\
\indent This paper is arranged as follows: 
in Section~\ref{3g}, the basic information of the third generation detectors are introduced. 
In Section~\ref{sec:method}, we present the information of sources in simulations and explain the Fisher matrix method used to estimate the localisation errors. And the results of the simulations are discussed in Section \ref{result_sec}. 
Finally, in Section \ref{final}, a discussion and a conclusion are provided.
\section{Third generation detectors}\label{3g}
\begin{figure}[h]
\centering
\includegraphics[width=8cm]{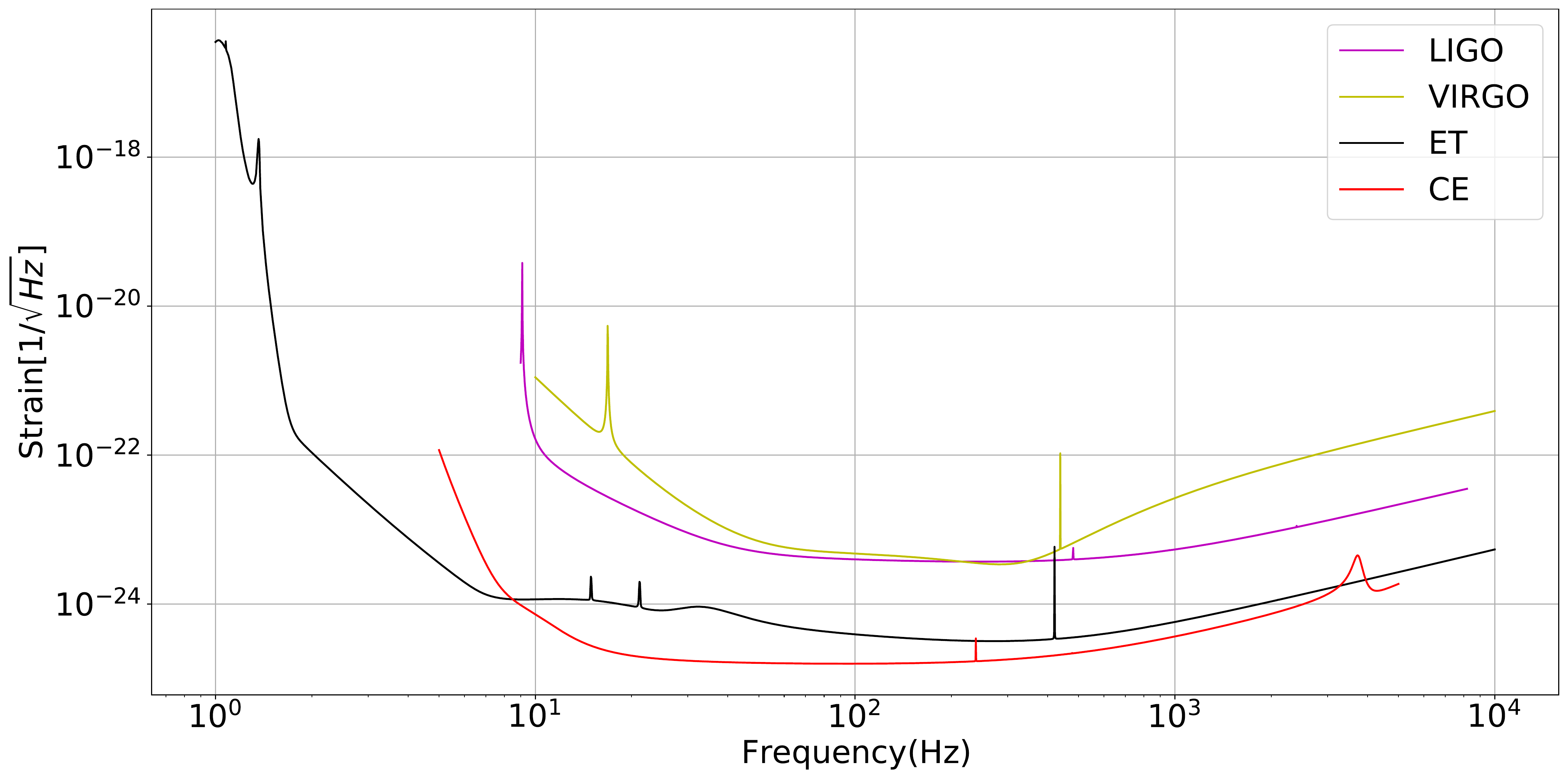}
\caption{The sensitivity curve of the second generation detectors LIGO, VIGO and the third generation detectors \ac{ET} and \ac{CE}. }\label{strain}
\end{figure}

\indent Third generation \ac{GW} detectors aim to improve the sensitivity 
across the entire frequency band.~\citep{2017CQGra..34d4001A, 2010CQGra..27s4002P}.  At present, the two proposed third generation detectors are the \ac{ET} and \ac{CE}.
\ac{ET} is composed of three interferometers with 10 km long arms. 
The interferometers will each have an opening angle equal to $60^{\circ}$ and each of them will be rotated relative to the others by $120^{\circ}$, forming 
an equilateral triangular structure. Proposed locations for \ac{ET} include Sardinia, Italy and the border region between the Netherlands, Belgium and Germany.
The target sensitivity of the \ac{ET} is shown in Figure~\ref{strain}. 
For frequency larger than 10Hz, compared to the Advanced LIGO and Advanced Virgo,
the sensitivity of \ac{ET} will be improved by a factor of $\sim 10$,
and even better for frequencies below $10$Hz.
A detailed description of ET can be found in \citet{2014ASSL..404..333P, 2010CQGra..27s4002P}. 

Unlike the \ac{ET}, \ac{CE} will have the typical L-shaped configuration employed by second generation detectors, with an arm length of 40 km. For frequencies $\geq 10$Hz, as shown in Figure \ref{strain},
the sensitivity is improved by a factor of $\sim 30$ compared
to Advanced LIGO and Advanced Virgo~\citep{2017CQGra..34d4001A, 2015PhRvD..91h2001D}.
The planned location for \ac{CE} is North America.
More extensive technology details about CE can be found in \citet{2019BAAS...51c.141R, 2019BAAS...51g..35R}.

With the improvement of sensitivity, \ac{ET} is expected to be able to detect mergers of \ac{BNS} at redshifts of about $2$, and it will be even higher for \ac{CE}~\citep{2018SSPMA..48g9805Z}. On the one hand, the improved sensitivity of \ac{GW} detectors at frequencies $\geq 10$Hz will directly lead to a higher \ac{SNR} so as to improve the localization of \ac{GW} events.  On the other hand, the improved sensitivity for the frequency band of $1$Hz - $10$Hz will 
greatly extend the in-band duration of \ac{GW} signals, so as to improve the localization as shown in \cite{2018PhRvD..97l3014C, 2018SSPMA..48g9805Z}.\\
\indent Assuming the detector is at the center or a spherical coordinate system, the antenna pattern of each interferometer of a \ac{ET}-like detector can be expressed as follows~
\citep{2012PhRvD..86l2001R}:
\begin{equation}
\centering
\begin{array}{r}F_{+}^{1}(\theta, \phi, \psi)=-\frac{\sqrt{3}}{4}\left[\left(1+\cos ^{2} \theta\right) \sin 2 \phi \cos 2 \psi\right. \\ +2 \cos \theta \cos 2 \phi \sin 2 \psi], \\ F_{\times}^{1}(\theta, \phi, \psi)=\frac{\sqrt{3}}{4}\left[\left(1+\cos ^{2} \theta\right) \sin 2 \phi \sin 2 \psi\right. \\ -2 \cos \theta \cos 2 \phi \cos 2 \psi],\\
F_{+, \times}^{2}(\theta, \phi, \psi)=F_{+, \times}^{1}\left(\theta, \phi+\frac{2 \pi}{3}, \psi\right),\\ F_{+, \times}^{3}(\theta, \phi, \psi)=F_{+, \times}^{1}\left(\theta, \phi-\frac{2 \pi}{3}, \psi\right)\end{array}\label{et_ap}
\end{equation}
where $F_{+}^{n}$ is the plus polarisation response of the $n$-th interferometer
and $F_{\times}^{n}$ is the cross polarisation response. 
The antenna patterns are a function of the azimuthal angle, $\theta$, the polar angle, $\phi$ and the polarization angle, $\psi$, of the \ac{GW} source. 
The antenna pattern of \ac{CE} can be expressed as~\citep{2011CQGra..28l5023S}:
\begin{equation}
\begin{array}{r}F_{+}(\theta, \phi, \psi)=\frac{1}{2}\left(1+\cos ^{2} \theta\right) \cos 2 \phi \cos 2 \psi \\ -\cos \theta \sin 2 \phi \sin 2 \psi, \\ F_{\times}(\theta, \phi, \psi)=-\frac{1}{2}\left(1+\cos ^{2} \theta\right) \cos 2 \phi \sin 2 \psi \\ -\cos \theta \sin 2 \phi \cos 2 \psi.\end{array}\label{ce_ap}
\end{equation}
where, as before, $F_{+}$ is the plus polarisation response for CE and $F_{\times}$ is the cross polarisation response. In addition, better sensitivity of detector such as \ac{ET} in low frequency band enables longer in-band signal duration, generally, 1~Hz corresponds to about 5 days as explained in section~\ref{sec:method}. Therefore, the azimuthal angle and polar angle will change over time as the earth keeps rotating, which means the antenna pattern will also be time-dependent. Additionally, due to the long duration of signal, the signal will experience Doppler effect because the the detector moves relative to the signal, in terms of Doppler effect, \citet{2018PhRvD..97l3014C} suggests the Doppler effect is not important to the localization uncertainty estimation, while for completeness, we also included the Doppler effect in analysis. \\
\indent For our study, we assume that the ET is located at the Virgo site in Italy (longitude, latitude) =  ($10.4^{\circ}\text{E},~43.7^{\circ}\text{N}$), 
and \ac{CE} at the LIGO Hanford site in the United States (longitude $-119.41^{\circ}\text{E}$, latitude $46.45^{\circ}\text{N}$). 
Though it is likely that the final locations of \ac{CE} and the \ac{ET} will not be at these assumed locations, 
the impact of any change in location within the United States and Europe will be small since our study is looking at the results averaged across a population of simulated signals. 
Additionally, we consider a third detector located in Australia (longitude $115.87^{\circ}\text{E}$, latitude $-31.95^{\circ}\text{N}$) for our study. 
Specifically, we assume two different scenarios. In the first scenario, the detector located in Australia is a detector with sensitivity and configuration identical to those of the \ac{ET}, denoted by \ac{ET}-A, while in the other scenario, the detector will be similar to \ac{CE}, referred to as \ac{CE}-A.
To further explore the localisation capabilities of different combinations of third generation \ac{GW} detectors, two detectors, denoted by \ac{ET}-L and \ac{CE}-V, 
are introduced. Similarly, \ac{ET}-L indicates a \ac{ET}-like detector at the location of LIGO Hanford, and CE-V refers to a \ac{CE}-like detector at the Virgo site. 

We simulate 4 networks of \ac{GW} detectors. The simulated networks can be found in Table~\ref{GWD}.
\begin{table*}[htbp]
  \centering
  \begin{tabular}{cccccccc}
      \hline
      \hline
        &   Australia & LIGO Hanford & Italy \\
        &  ( $115.87^{\circ}\text{E}$,$-31.95^{\circ}\text{N}$)&  ($-119.41^{\circ}\text{E}$, $46.45^{\circ}\text{N}$)&($10.4^{\circ}\text{E}$,  $43.7^{\circ}\text{N}$) \\
       \hline 
       2ET1CE& ET-A & CE & ET  \\  
       
       3ET& ET-A & ET-L & ET \\  
       
       1ET2CE& CE-A & CE & ET \\  
      
    3CE& CE-A & CE & CE-V \\                
      \hline
       \hline
  \end{tabular}
  \caption{The simulated GW detector networks in this paper.
  The first row is the locations of the \ac{GW} detectors in longitude and latitude. 
  The left column shows the names of the \ac{GW} detector networks, which directly indicates the network configuration. 
  \ac{ET}-A and \ac{ET}-L represent an ET-like detector in Australia and the LIGO Hanford site respectively.
  Similarly, \ac{CE}-A and \ac{CE}-V indicate a CE-like detector located in Australia and Italy.\label{GWD}}
  \end{table*}
We use the numbers and the types of the detectors in a network to refer to the network. 
For example, 2\ac{ET}1\ac{CE} refers to a network consisting of $2$ \ac{ET}-like detectors and $1$ \ac{CE}-like detector, and 3\ac{ET} to a network consisting of $3$ \ac{ET}-like detectors. 
We estimate the localization uncertainty and the prospect for early warning for these networks.
\section{Methodology}\label{sec:method}
\subsection{Binary neutron star simulations}

Generally, the evolution of BNSs is considered to go through three stages: inspiral, merger and ringdown. The \ac{GW} signal during the adiabatic inspiral phase can be well described by post-Newtonian~\citep{2012PhRvD..86f9903A}. 
Our major concern is the localization performance achieved before the merger of BNSs to evaluate early warning capabilities, 
thus our demand for an accurate description of the merger and ringdown is not high. In this context,  we use the TaylorT3 waveform (specifically, Eq (3.10b) in~\cite{2009PhRvD..80h4043B}) approximant to simulate signals in this work. In addition to the signal templates,  the duration of simulated signals also need to be taken into consideration. The duration of simulated signal, in other words, the time remaining to merger,  can be simply estimated according to a leading order post-Newtonian approximation as follows~\citep{Maggiore}: 
 \\
\begin{equation}
\tau_{\mathrm{c}}=\frac{5}{256} \frac{c^{5}}{G^{\frac{5}{3}}} \frac{\left(\pi f_{\mathrm{s}}\right)^{-\frac{8}{3}}}{\mathcal{M}^{\frac{5}{3}}},\label{time_to_merger}
\end{equation}
where $\tau_{\mathrm{c}}$ is the time to merger, $c$ the speed of light, $G$ the gravitational constant, 
$\mathcal{M}$ the chirp mass ($\mathcal{M}=\frac{\left(m_{1} m_{2}\right)^{3 / 5}}{\left(m_{1}+m_{2}\right)^{1 / 5}}$, where $m1, m2$ are component masses), and $f_{\mathrm{s}}$ the low cut-off frequency.  It should be noted that component masses, chirp masses, and total masses ($M = m_1+m_2$) in our simulations are all redshifted to observed mass $M_{\mathrm{obs}}$ by:
\begin{equation}
M_{\mathrm{obs}}=M_{\mathrm{local}}(1+z).
\end{equation}
Now assuming a low cut-off frequency of 1~Hz,  and a \ac{BNS} of $1.4M_\odot$-$1.4M_\odot$,  the in-band duration of the \ac{GW} signal can be estimated to be about 130 hours. Naturally, during detection for signals with duration of 130 hours,  the earth keeps rotating,  in this context, time-dependent antenna pattern needs to be applied for a reliable estimation for localization. \\
\indent To estimate the localisation for different populations of \ac{BNS} mergers, two scenarios are considered,  \acp{BNS} at fixed distances, and a population of \acp{BNS} with a redshift distribution following the power law \ac{DTD} described in~\citet{2019ApJ...878L..13S}.  
For \acp{BNS} at fixed distances, we select $40\text{Mpc}$, $200\text{Mpc}$, $400\text{Mpc}$, $800\text{Mpc}$ and $1600\text{Mpc}$ as examples. 
Specifically, 100 \acp{BNS} are simulated for each distance. 
$500$ \acp{BNS} sources following the \ac{DTD} of redshift up to $2$ are simulated.  

\begin{figure}[h]
\centerline{
\includegraphics[width=10.2cm]{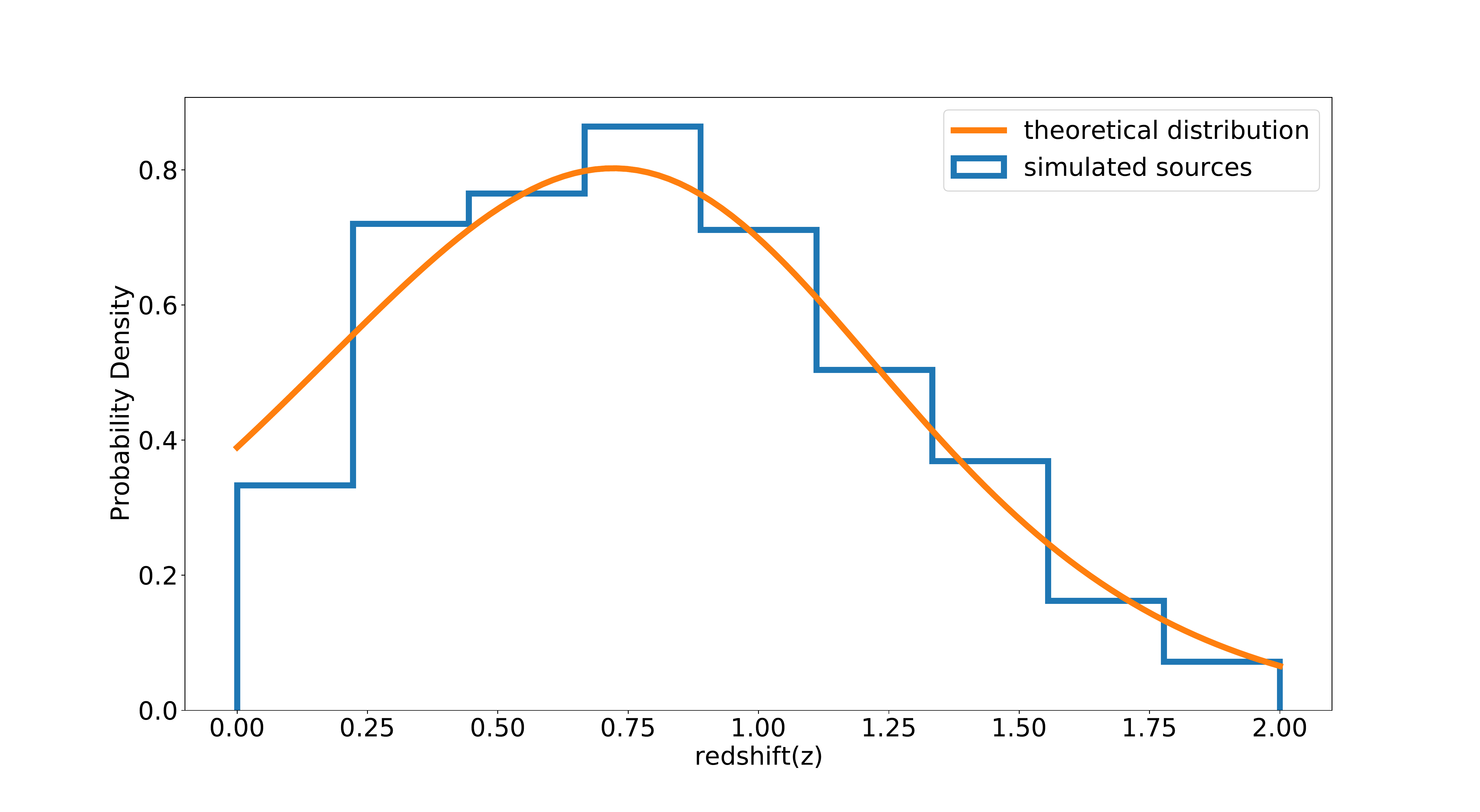}}
\caption{The redshift distribution of simulated sources.  
The x-axis represents the redshift. The y-axis represents the normalized distribution which makes the area under the curve is 1.  
The orange line is the redshift distribution calculated using Eq. \ref{ndot} with $t_\text{{min}} = 1 \text{Gyr}, \Gamma = -1.5$,  and the blue line is the distribution of simulated sources.}\label{injection}
\end{figure}

The mass of the simulated BNSs are all $1.4M_{\odot}$-$1.4M_{\odot}$ in local frame.
The sky locations (the right ascension $\alpha$ and declination $\delta$) of the simulated \ac{BNS} mergers for all cases are randomly sampled to simulate uniform distribution in the whole sky.
The cosine of inclination angle $cos~\iota$ are randomised in range between -1 and 1,  polarization angle $\psi$ are also randomised in range between 0 and $2\pi$. \\
\indent For the \ac{BNS} mergers following the \ac{DTD} used in this work, the \ac{BNS} merger rate as a function of redshift $z$ can be expressed as,
 \begin{equation}
R(z)=\int_{z_{b}=10}^{z_{b}=z} \lambda \frac{d P_{m}}{d t}\left(t-t_{b}-t_{\min }\right) SFR\left(z_{b}\right) \frac{d t}{d z}\left(z_{b}\right) d z_{b}. \label{ndot}
\end{equation}
Here, $R(z)$ represents the BNS intrinsic merger rate density ($\text{Mpc}^{-3}\text{year}^{-1}$), and $SFR$ represents cosmic star formation rate density ($\text{M}_{\odot}\text{yr}^{-1}\text{Mpc}^{-3}$). $\lambda$ is the \ac{BNS} mass efficiency which is assumed to be a constant $10^{-5} \mathrm{M}_{\odot}^{-1}$, $t_b$ is the time at the redshift $z_b$ and $d P_{m} / d t$ is the \ac{DTD}, 
which is assumed to follow a power law distribution with a minimum delay time, and $\Gamma$ as the power law index, given by
 \begin{equation}
d P_{m} / d t  \propto t^{\Gamma},\label{dpm_dt}
\end{equation}
$t_{\text{min}}$ in Eq~(\ref{ndot}) is the minimum delay time which corresponds to the time since the star first joins the main sequence on the Hertzsprung-Russell diagram. 
In this paper, we adopt $t_\text{{min}} = 1$Gyr and $\Gamma = -1.5$. 
The derivative of time, $dt/dz$, with respect to redshift is given by
\begin{equation}
d t / d z=-\left[(1+z) E(z) H_{0}\right]^{-1},
\end{equation}
where 
\begin{equation}
E(z) = \sqrt{\Omega_{m,0}(1+z)^{3}+\Omega_{k,0}(1+z)^{2}+\Omega_{\Lambda}(z)},
\end{equation}
where $\Omega_M$ is the matter density, $\Omega_{\Lambda}$ 
vacuum density , $\Omega_K$ curvature, and  $H_0$ the Hubble constant.
We adopt the values from the Planck 2015 results~\citep{2016A&A...594A..13P}.
In particular, $\Omega_M = 0.308$, $\Omega_{\Lambda} = 0.692$, $\Omega_K = 0$, and  $H_0 = 67.8 ~\mathrm{km  ~s^{-1}~Mpc^{-1}}$ respectively.
For the cosmic star formation rate density $SFR\left(z\right)$, we adopted the following formula from~\citep{2019ApJ...878L..13S, 2014ARA&A..52..415M}:
\begin{equation}
SFR(z)=0.015 \frac{(1+z)^{2.7}}{1+[(1+z) / 2.9]^{5.6}} \mathrm{M}_{\odot} \mathrm{yr}^{-1} \mathrm{Mpc}^{-3},
\end{equation}
where $z$ represents the redshift. In Eq.~\ref{ndot} we chose the lower limit of integration to be redshift of 10 based on \citet{2019ApJ...878L..13S},  because the choice of maximum redshift has little impact on the calculation. 
In the case of $t_\text{{min}}$ = 1 Gyr, $\Gamma = -1.5$, we calculated its corresponding \ac{BNS} merger rate density for a redshift range from 0 to 2 via Eq.~\ref{ndot}.  And by assuming the integral of \ac{BNS} merger rate density over redshift range from 0 to 2 is 1 (that is, all sources are distributed within redshift range of 0 - 2),  the redshift distribution (namely, the probability density function of redshift) of sources could be obtained as shown in Figure~\ref{injection}.  We simply chose 2 as maximum redshift cut-off.  \\

\subsection{Fisher matrix analysis}
The Fisher matrix translates errors on observed quantities measured directly into constraints on parameters of interest in the underlying model~\citep{2011IJMPD..20.2559B}, which has tested as an applicable tool in a series of researches ~\citep{2009NJPh...11l3006F, 2010PhRvD..81h2001W, 2011CQGra..28j5021F, 2018PhRvD..97l3014C, 2018PhRvD..97f4031Z, 2018PhRvD..97j4064M} to estimate localization errors for \ac{GW} detectors with not high computation cost compared to other methods.  As elaborated in \citet{2008PhRvD..77d2001V},  the inverse Fisher matrix represents the covariance matrix of parameter errors for the true signal, which could also be interpreted as the frequentist error covariance for the maximum-likelihood parameter estimator assuming Gaussian noise in the case of high SNR limit (that means, the waveforms could be considered as linear functions of source parameters).  And in the high SNR limit,  the maximum-likelihood naturally achieves the Cramér-Rao bound. Therefore that's why we set a SNR threshold for further localization uncertainty analysis. Only those signals detected with SNR larger than or equal to SNR threshold are seen as detectable signals. In this paper, we chose a network SNR of 8 as detectable threshold. In addition, for waveforms of low SNR or waveforms with poor priors estimation for its parameters, the Fisher matrix could not be applied to parameter estimation of \ac{GW} data credibly. Also, it's worth noting that the detector network consisting of three detectors significantly improves the sky localization accuracy compared to two-detector networks and thus enhances the reliability to estimate localization error via the Fisher matrix method \citep{2015ApJ...804..114B}.\\
\indent We consider a vector, $\boldsymbol{\theta}$, of $9$ parameters in our Fisher matrix analysis as in \citet{2018PhRvD..97l3014C}: 
$\alpha$; $\delta$; the arrival time of the signal at the center of the earth, $t_0$; the log of distance, $log_{10}d$; polarization angle, 
$\psi$; the log of the total mass, $log_{10}M$; the cosine of the inclination angle, $cos~\iota$;
the symmetric mass ratio, $\eta = m_1 \times m_2/M^2$; 
the initial phase of the wave when it arrives at the center of the earth, $\phi_0$.
The elements of $\boldsymbol{\theta}$ in the Fisher matrix then could be obtained by,
\begin{equation}\label{eq:FIM}
\rm FIM_{ i j}=\sum_{d=1}^{N} 2 \int_{0}^{\infty} \frac{\frac{\partial h_{d}^{*}}
{\partial \theta_{i}} \frac{\partial h_{d}}{\partial \theta_{j}}+\frac{\partial h_{d}^{*}}{\partial \theta_{j}} \frac{\partial h_{d}}{\partial \theta_{i}}}{P_{d}} d f,
\end{equation}
where $h_{\text{d}}$ represents the incoming gravitational wave strain in the frequency domain at the $\text{d}^{th}$ 
detector. $\frac{\partial h_{\text{d}}}{\partial \theta_{i}}$ 
denotes the partial derivative of $h_{\text{d}}$ with respect to the $i^{th}$ unknown parameter $\theta_{i}$.  
$P_\text{d}$ refers to the power spectral density for $\text{d}^{th}$ interferometer.
The optimal \ac{SNR}, $\rho_\text{d}$, for the $\text{d}^{th}$ detector can be expressed as,
\begin{equation}
\rho_\text{d}=\sqrt{4 \int_{0}^{\infty} \frac{\left|h _\text{{d}}(\boldsymbol{\theta}, f)\right|^{2}}{P_\text{d}(f)} d f}.
\end{equation}
The combined \ac{SNR} for a signal from a network of detectors can be obtained by
\begin{equation}
\rho_\mathrm{network} = \sqrt{  \sum_{\text{d}=1}^{N}\rho_\text{d}^{2} }.
\end{equation} 
In simulation, we divide the strain data in the time domain into S 
segments with length of 100 seconds (the last segment shorter than 100 seconds), 
and use Eq. \ref{eq:FIM} to calculate the Fisher matrix for each segment. 
The Fisher matrix for the entire signal is then obtained by summing the Fisher matrices over the segments,
\begin{equation}
 \text{FIM}_{i j}^{\text{t}}=\sum_{k=1}^{S} \text{FIM}_{i j}^{k},
\end{equation}
where $\text{t}$ represents variable for the entire signal, and $k$ indicates the $k^{\text{th}}$ segment.
The total accumulated \ac{SNR} of single detector for the entire signal $\rho_d^t$
is obtained by 
\begin{equation}
\rho_d^\text{t}=\sqrt{\sum_{k=1}^{S}\left(\rho_d^{k}\right)^{2}},
\end{equation}
where $\rho_d^k$ represents \ac{SNR} of $k^{\text{th}}$ segment for  $d^{\text{th}}$ detector.
The covariance matrix can then be derived by computing the inverse of the Fisher matrix,
\begin{equation}
{\text{cov}_{i j}^\text{t}={(\text{FIM}_{i j}^\text{t})}^{-1}.}
\end{equation}
The localization uncertainty $\Delta \Omega$ 
is given by,
\begin{equation}
\Delta \Omega= \pi \sqrt{\lambda_{\alpha} \lambda_{\delta}} \cos \delta,
\end{equation}
where $\lambda_{\alpha}$ and $\lambda_{\delta}$ are the eigenvalues of $\mathrm{cov}^t_{i j}$ corresponding to $\alpha$ and $\delta$. 
Finally, the localization uncertainty at confidence level $p$ is obtained by,
\begin{equation}
\Delta \Omega_{p}=-2 \log (1-p) \Delta \Omega.\label{eq_loc}
\end{equation}

\section{Results}\label{result_sec}
We use a network SNR of $8$ to determine whether an event is detectable.
To avoid ambiguity, the localization errors shown in the remaining of this paper are all at  $90\%$ confidence. 
\subsection{Sources at fixed distances}
\subsubsection{Localization uncertainty}
Almost all \ac{BNS} sources simulated at the chosen fixed distances were observed with network SNR greater than 8. In Table~\ref{snr_er_number}, we list the percentage of the detectable sources which can be localised to within $30\mathrm{deg^2}$, $10\text{deg}^2$, $5\text{deg}^2$, and $1\mathrm{deg^2}$ with each detector network. 

All simulated BNS signals at 40 Mpc are detectable to all simulated detector networks and can be localised to an area smaller than $1\mathrm{deg^2}$ and are, thus, not shown in Table~\ref{snr_er_number}. We also note that all sources at 200 Mpc are localized to within $1\text{deg}^2$ by all networks. As expected, the localization uncertainty gets increases with distance increases.  Less than 10\% sources at 1600 Mpc could be detected within $1\text{deg}^2$ by the 3ET network and 2ET1CE networks, while 1ET2CE and 3CE can localize respectively 16\% and 40\% sources.

\begin{table}[htbp]
\centering
  \begin{tabular}{cccccccc}
      \hline
      \hline
       &  & 3ET & 2ET1CE & 1ET2CE & 3CE \\
       \hline
    \multirow{4}{1.5cm}{200Mpc}& $30\mathrm{deg^2}$ & 100\% & 100\% & 100\% & 100\% \\
       & $10\mathrm{deg^2}$ & 100\% & 100\% & 100\% & 100\%\\
       & $5\mathrm{deg^2}$ & 100\% & 100\% & 100\% & 100\%  \\
        & $1\mathrm{deg^2}$ & 100\% & 100\% & 100\% & 100\% \\

       \hline
       
    \multirow{4}{1.5cm}{400Mpc}& $30\mathrm{deg^2}$ & 100\% & 100\% & 100\% & 100\% \\
       & $10\mathrm{deg^2}$ & 100\% & 100\% & 100\% & 100\%\\
       & $5\mathrm{deg^2}$ & 100\%& 100\% & 100\% & 100\% \\
        & $1\mathrm{deg^2}$ & 91\%& 89\% & 95\% & 96\%  \\
       
       \hline
       
     \multirow{4}{1.5cm}{800Mpc}& $30deg^2$ & 100\% & 100\% & 100\% & 100\% \\
       & $10\mathrm{deg^2}$ & 100\% &  99\% & 98\% & 98\%\\
       & $5\mathrm{deg^2}$ & 95\% & 93\% & 96\% & 96\%  \\
        & $1\mathrm{deg^2}$ & 50\%& 53\% & 64\% & 77\%  \\

       \hline
      \multirow{4}{1.5cm}{1600Mpc}& $30\mathrm{deg^2}$ & 99\% & 99\% & 98\% &98\% \\
       & $10\mathrm{deg^2}$ &80\% & 79\% & 89\% & 91\%\\
       & $5\mathrm{deg^2}$ & 59\%& 58\% & 70\% & 82\% \\
        & $1\mathrm{deg^2}$ & 7\% & 9\% & 16\% & 40\%  \\
      \hline
      \hline
  \end{tabular}
  \caption{A table showing the fraction of detectable sources which can be localised to within $\mathrm{30deg^2, 10deg^2, 5deg^2}$, and $1\text{deg}^2$ with the \ac{GW} networks.}\label{snr_er_number}
\end{table}

\begin{table}[htbp]
  \centering
  \begin{tabular}{p{1.6cm}p{1cm}p{0.9cm}p{1cm}p{1.3cm}p{0.76cm}}
      \hline
      \hline
       &  & 3ET ($\text{deg}^2$) & 2ET1CE ($\text{deg}^2$) & 1ET2CE ($\text{deg}^2$) & 3CE ($\text{deg}^2$) \\
      \hline
        \multirow{3}{1.5cm}{200Mpc}& 90\% & $0.23$ & $0.25$ & $0.18$ & $0.14$ \\
       & 50\% & $0.06$ & $0.06$ & $0.04$ & $0.02$  \\
       & 10\% & $0.02$& $0.02$ & $0.01$ & $0.01$  \\
       \hline
       
       \multirow{3}{1.5cm}{400Mpc}& 90\% & $0.94$ & $1.02$ & $0.72$ & $0.56$ \\
       & 50\% & $0.23$ & $0.22$ & $0.15$ & $0.09$ \\
       & 10\% &$0.07$& $0.06$ & $0.06$ & $0.02$  \\
       \hline
       
      \multirow{3}{1.5cm}{800Mpc} & 90\% & $3.75$& $4.07$ & $2.89$ & $2.25$\\
       & 50\% & $0.93$ & $0.91$ & $0.60$ & $0.35$  \\
       & 10\% & $0.26$& $0.26$ & $0.23$ & $0.09$  \\
       \hline
       \multirow{3}{1.5cm}{1600Mpc}& 90\% & $14.58$ &$14.65$ & $10.07$ & $9.01$\\
       & 50\% & $3.69$ & $3.57$ & $2.30$ & $1.39$  \\
       & 10\% & $1.06$& $1.03$ & $0.90$& $0.38$  \\
      \hline
      \hline
  \end{tabular}
  \caption{The 90\% credible regions of localization uncertainty of every detector network for binary neutron star mergers respectively at 200 Mpc,  400 Mpc,  800 Mpc,  1600 Mpc with SNR$>$8. $90\%$, $50\%$ and $10\%$ respectively represents the best localized 100\%, 90\% and 50\% of the detectable sources. 
\\}\label{loct}
  \end{table}

\begin{figure*}[h]
\begin{minipage}[t]{1\textwidth}
\centering
\includegraphics[width=15cm]{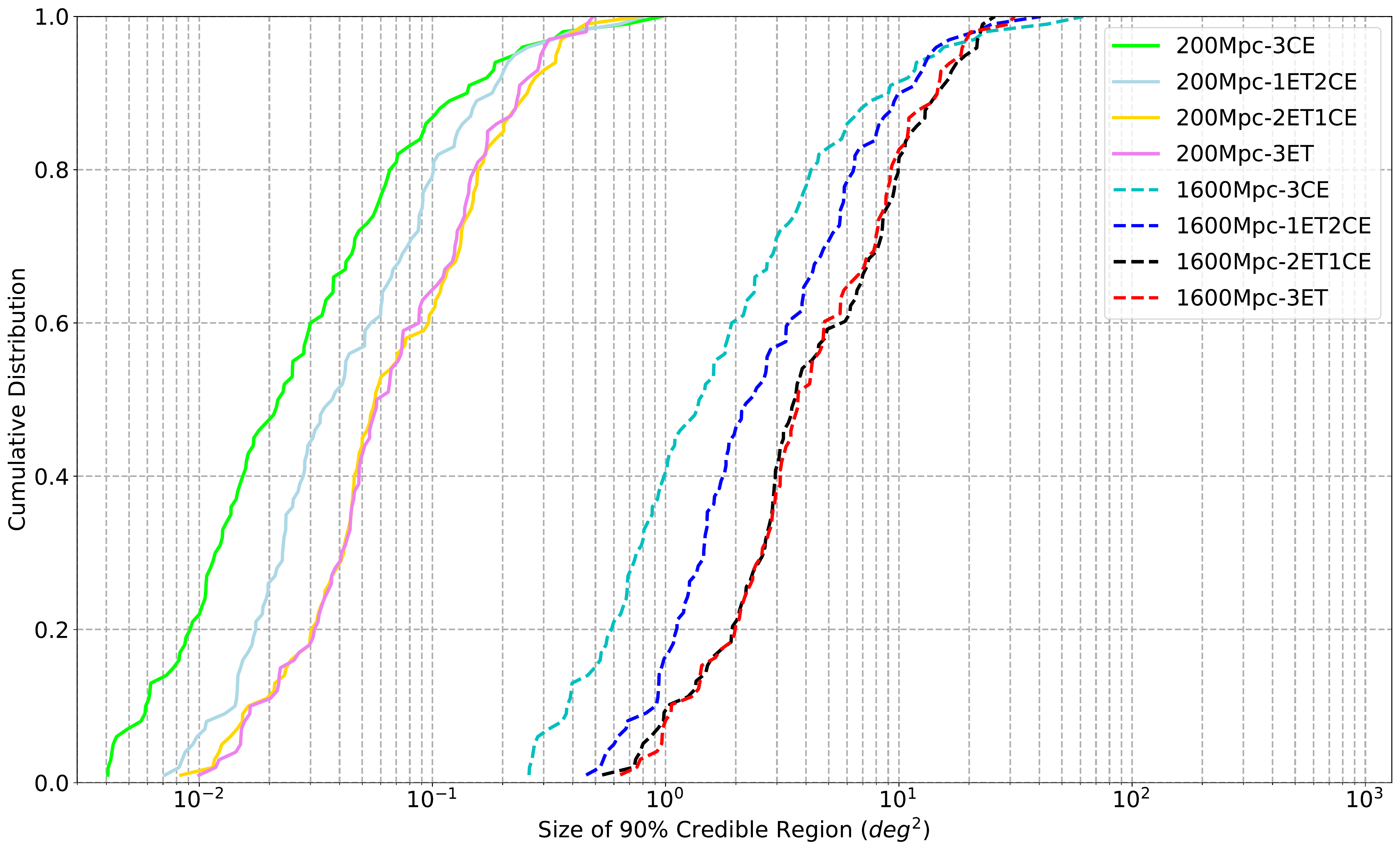}
\caption{The cumulative distributions of the size of the $90\%$ credible regions for the \ac{BNS} mergers at 200 Mpc and at 1600 Mpc.  The x-axes show the size of the 90\% credible region and the upper limit of the x-axes corresponds to the size of the whole sky. }\label{er_200_1600}
\end{minipage}

\subfigure[$1\text{deg}^2$] { \label{fig:a} 
\includegraphics[width=1\columnwidth]{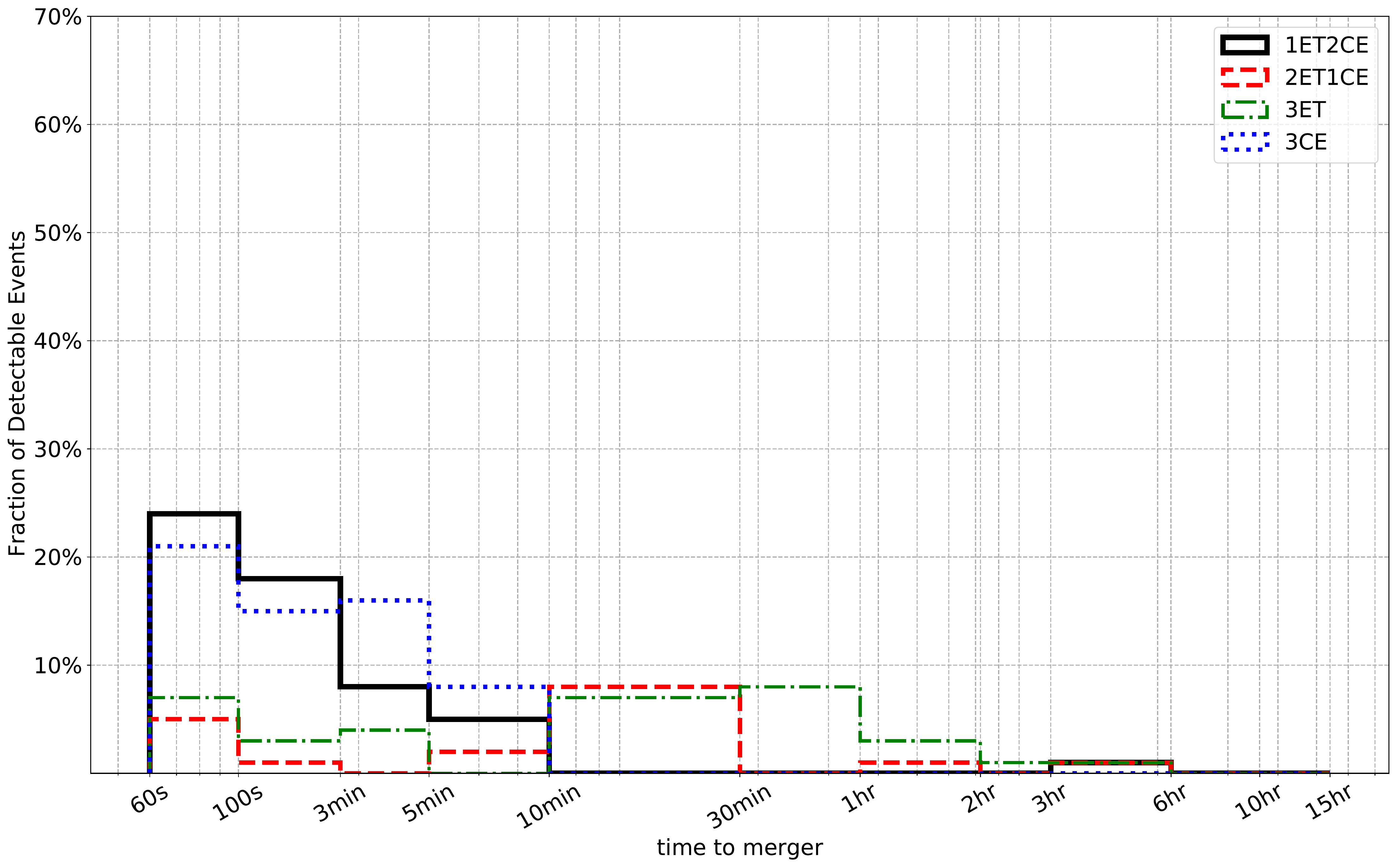}} 
\subfigure[$5\text{deg}^2$] { \label{fig:b} 
\includegraphics[width=1\columnwidth]{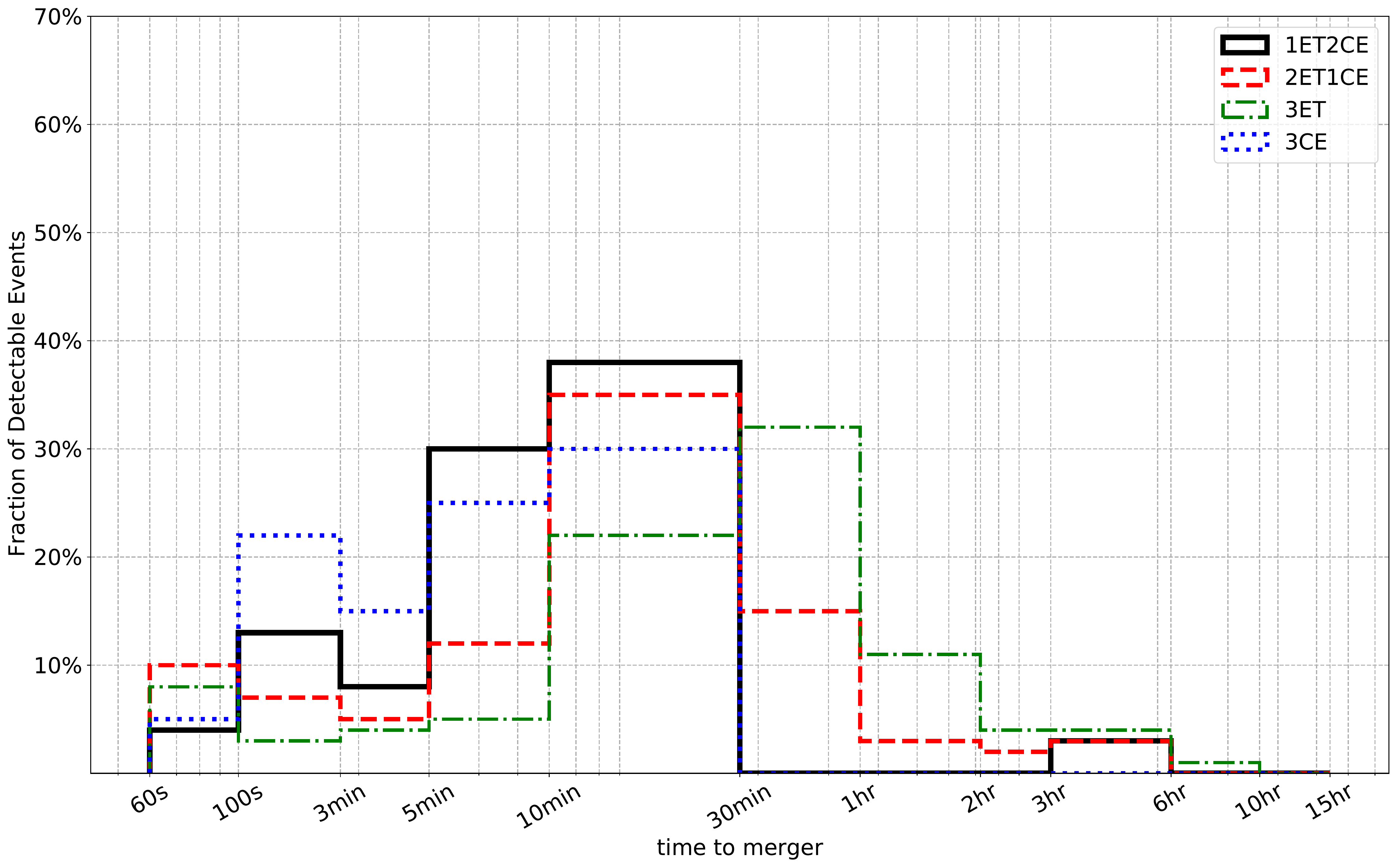}} 

\subfigure[$10\text{deg}^2$] { \label{fig:c} 
\includegraphics[width=1\columnwidth]{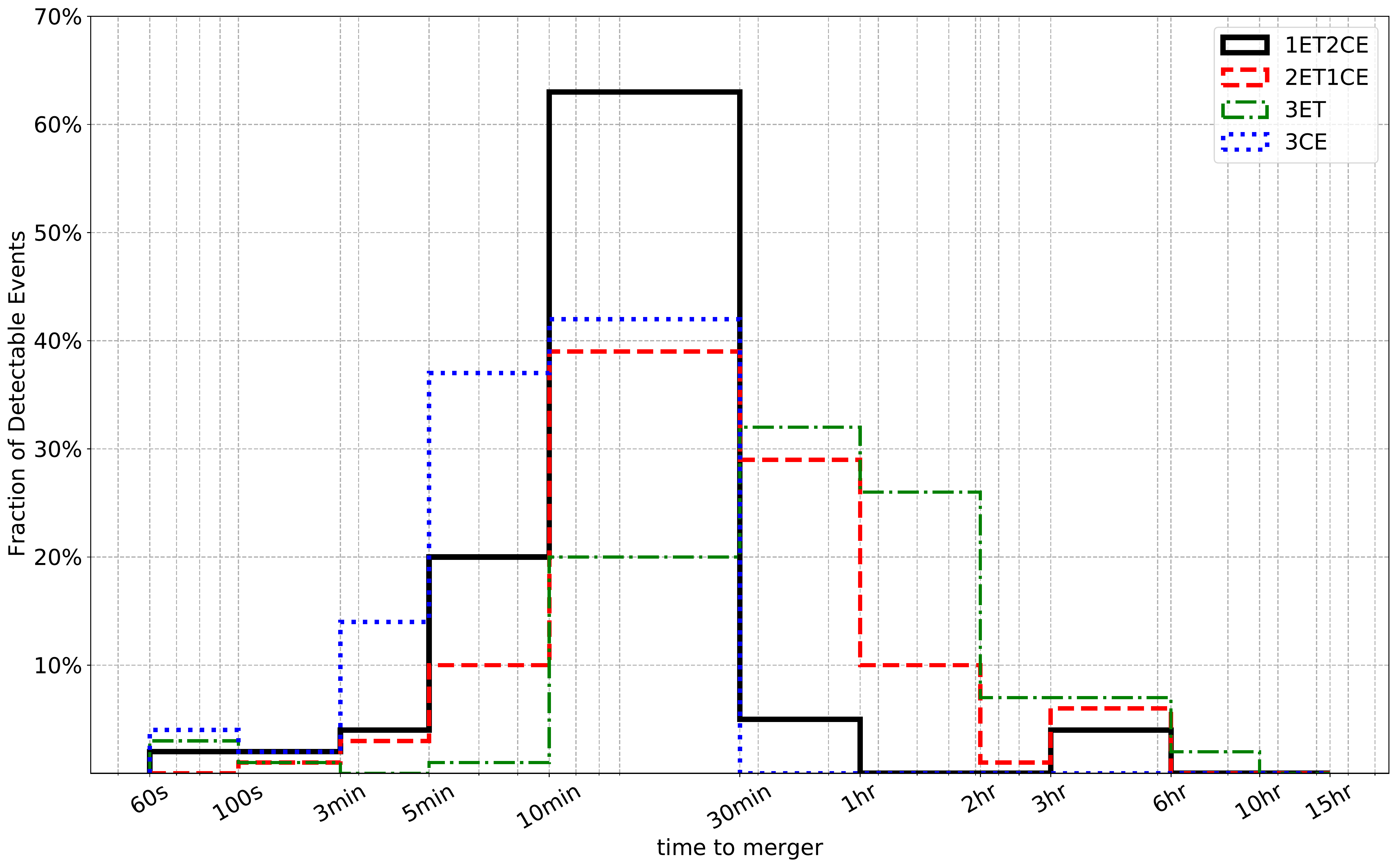}} 
\subfigure[$30\text{deg}^2$] { \label{fig:d} 
\includegraphics[width=1\columnwidth]{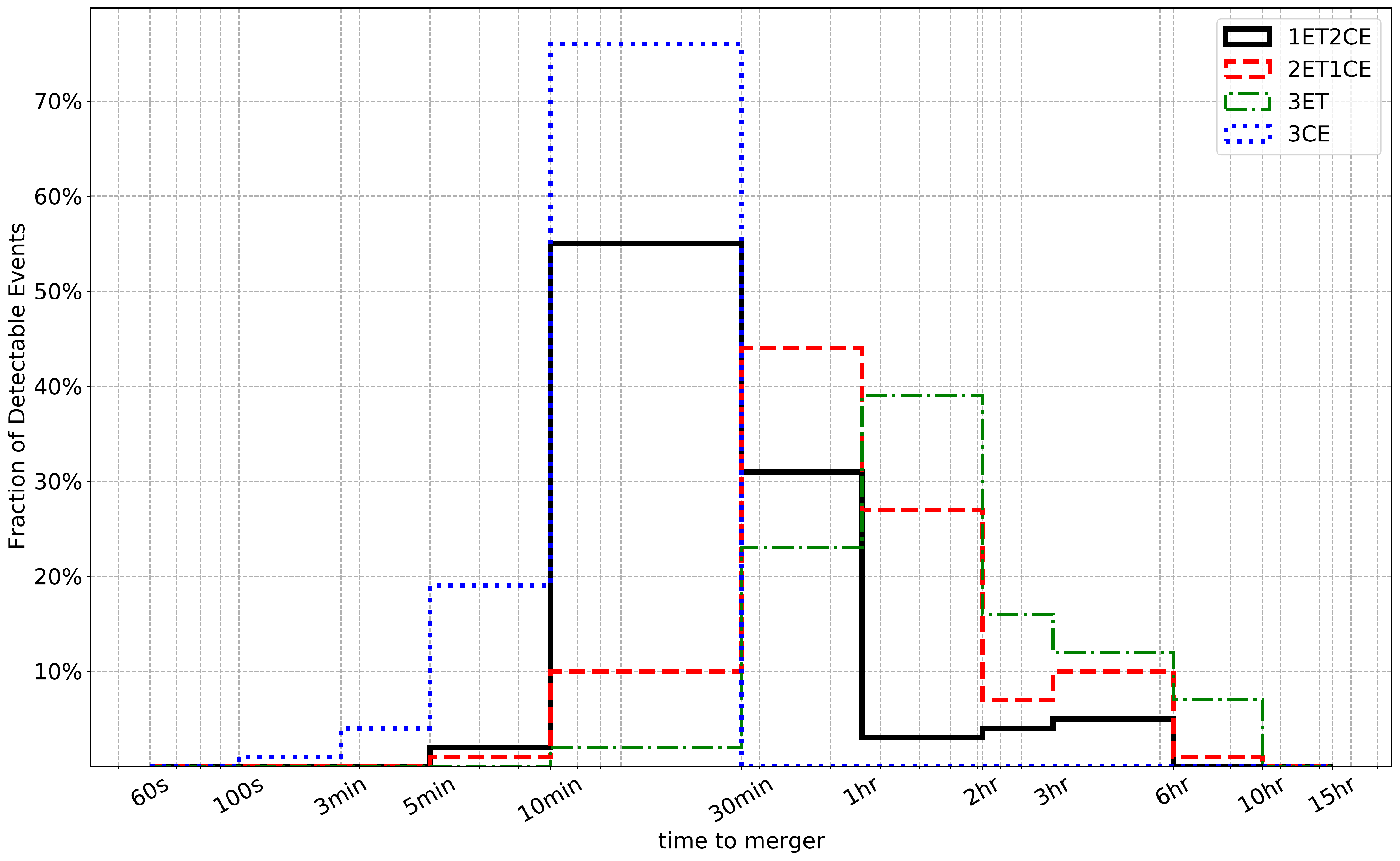}}
\caption{Histograms showing the distributions of time to merger for the \ac{BNS} mergers at $200$ Mpc with 4 detector networks.  The x-axis is the time to merger when the signal meets the early warning criteria. The y-axis is the fraction of detectable events that achieve these early warning criteria.  The percentage of sources that could be detected within required localization uncertainty could be found in Table~\ref{snr_er_number}.}\label{200EW}
\end{figure*}

\indent The cumulative distribution of the size of the 90\% credible regions for the \ac{BNS} mergers at fixed distances are presented in Figure~\ref{er_200_1600}. Results for 200 Mpc and 1600 Mpc are shown as an example for simplicity here and results of other fixed distances are presented in Appendix.  We note that the relative behaviour of the 4 networks remain largely unchanged for sources at 200 Mpc and 1600 Mpc. Taking 200 Mpc as an example, there's a trend of gradual change from 3CE to 1ET2CE to 2ET1CE network which gradually accumulate more higher values of localization uncertainty. Additionally, the cumulative distribution of 3ET network is very similar to that of 2ET1CE network.\\
\indent In Table~\ref{loct}, we present the upper limits of the size of the $90\%$ credible regions for the $90\%$, $50\%$ and $10\%$ best localized sources. 
Particularly, $90\%$ of the best localized sources at 200 Mpc can be constrained to within an area of $\sim0.3 \text{deg}^2$ with all 4 networks. 
Even for sources at a distance equal to $1600$Mpc, the size of the $90\%$ credible region is still smaller than $15\text{deg}^2$, indicating a promising prospect for \ac{EM} follow-up observations.
Furthermore, 50\% of the best localized sources respectively at 200 Mpc, 400 Mpc, 800 Mpc, 1600 Mpc,  could be limited to within $0.07 \text{deg}^2$, $0.30\text{deg}^2$, $1.00 \text{deg}^2$, $4.00 \text{deg}^2$ by all GW networks.  
We also found that the localization uncertainty achieved by 2ET1CE for 90\% best localized sources at any individual distance is $\sim1.4$ times of that of 1ET2CE network. And it becomes $\sim 1.5$ times for 50\% best localized sources. For 10\% best localized sources, the ratio ranges from 1 to 1.18. Overall, as shown in this table, 2ET1CE network has localization uncertainty no less than 1ET2CE network for simulated sources at fixed distances. 

\subsubsection{Early warning capabilities}
As mentioned in Section \ref{sec:method}, the in-band duration of the \ac{GW} signals from \ac{BNS} mergers have been significantly extended for third generation detectors. It is therefore possible for a signal to accumulate enough 
\ac{SNR} to be deemed significant before the merger.  This in turn makes it possible that a trigger may be released prior to merger, increasing the likelihood of a successful \ac{EM} follow-up observation.

To evaluate the early warning prospect of the simulated networks of \ac{GW} detectors, we require two conditions to be satisfied before a trigger can be released.
The first condition is that the signal has to accumulate a network \ac{SNR} of no less than $8$ at the time the trigger is released, and the other condition is that
the source of the signal has to be localised to an error no greater than a given certain area with $90\%$ confidence.  The second condition is required because a signal may accumulate enough \ac{SNR} to the point where the event is considered significant prior to merger,  but the localisation error is still too large for any meaningful \ac{EM} follow-up observations to be carried out. These two conditions will be referred to as early warning criteria in the remaining of this paper.  Furthermore, in terms of the second condition of early warning, we chose localization area of respectively $30~\text{deg}^2$,  $10~\text{deg}^2$, $5~\text{deg}^2$ and $1~\text{deg}^2$ as requirements to evaluate corresponding time to merger distribution under these conditions. \\
\indent Figure~\ref{200EW} presents the histogram distribution of time to merger of 4 detector network for BNSs at 200Mpc,  where each subplot denotes the time to merger distribution given maximum allowable localization uncertainty of respectively
$\mathrm{30~deg^2}$, $10~\mathrm{deg^2}$, $5~\mathrm{deg^2}$ and $1~\mathrm{deg^2}$.  It should be noted that the sum of percentage
for each network may not be 100\% because there are some sources could be detected within the required localization uncertainty only at the time of merger. 
Early warning of BNSs at 200Mpc will be discussed in detail here,
the related figures of BNSs at other fixed distances could be found in Appendix. \\
\indent Given the maximum localization uncertainty of $30\mathrm{deg^2}$, 100\% BNS signals at 200 Mpc could be detected by all tested GW detector networks at least 100 seconds before merger.  The time to merger of BNSs detected by 1ET2CE network ranges from 5 minutes to 6 hours, and is concentrated in range of 10 minutes to 1 hour.  While for 2ET1CE network, it ranges from 5 minutes to 10 hours, and is mainly distributed between 10 minutes and 6 hours.  The range of time to merger indicates a significant possibility of third generation \ac{GW} detector networks to give early warning alert to \ac{EM} follow-up observations.\\
\indent When the maximum allowable localization uncertainty decreases, to $10\mathrm{deg^2}$, $5\mathrm{deg^2}$ and $1\mathrm{deg^2}$, detectable sources which meet the early warning criteria become less and less, part of BNS signals could only meet the early warning criteria at the time of merger, and the distribution of time to merger move towards to a direction of smaller value. \\
\subsection{Astrophysical population}
\subsubsection{Signal-to-noise-ratio distributions}\label{snr_dtd}
\begin{figure}[h]
\centerline{
\includegraphics[width=1.2\columnwidth]{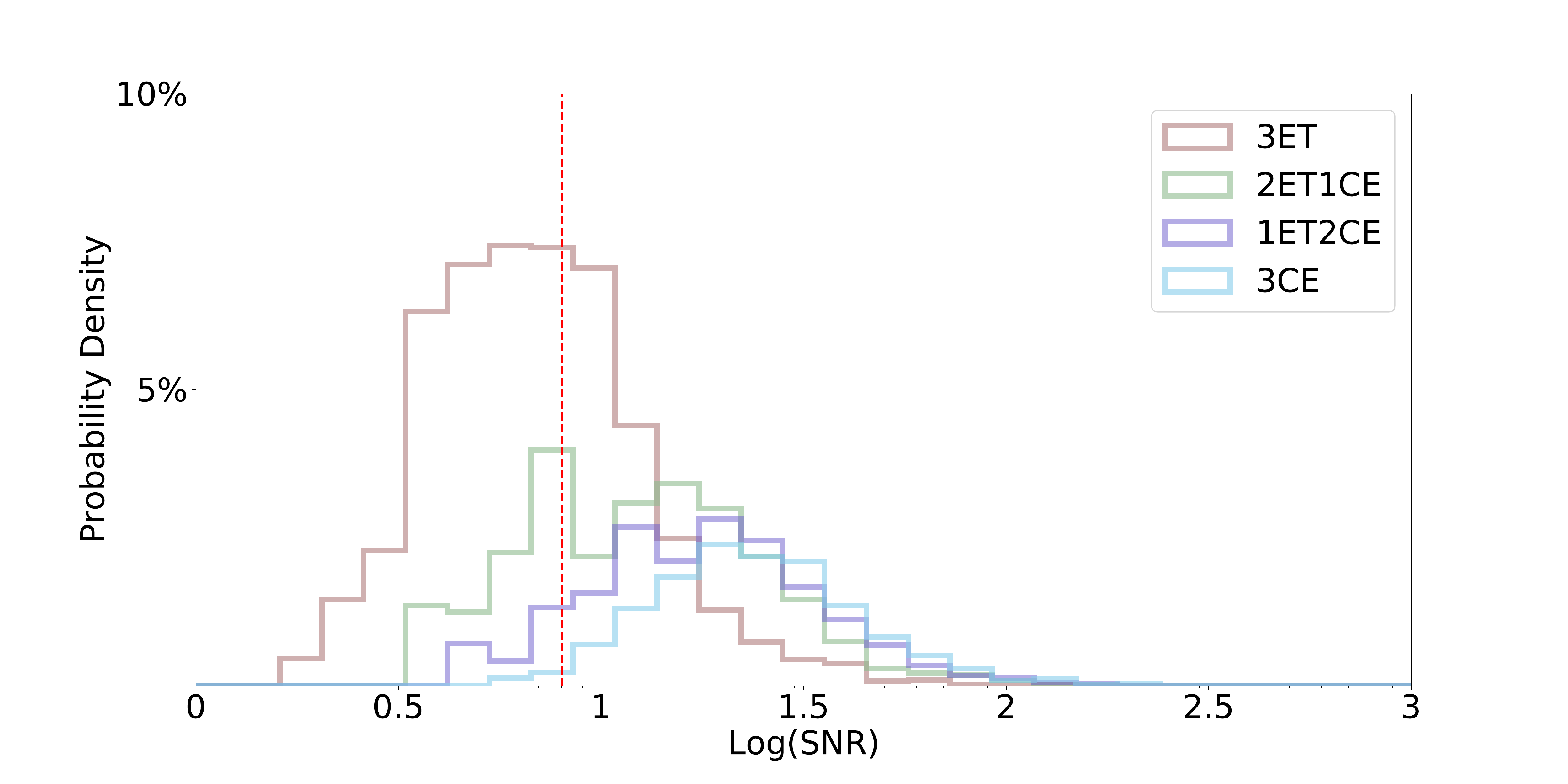}}
\caption{The probability density distributions of the \ac{SNR} for the \ac{BNS} which follow the \ac{DTD}. The horizontal axis represents the \ac{SNR} in log scale. 
The network combined SNR threshold of 8 is denoted by a red dashed vertical line.  
The area under the histogram is normalized to be 1. 
}\label{SNR_hist_3G}
\end{figure}

For the \ac{BNS} mergers following the \ac{DTD}, the SNR distribution is presented in Figure~\ref{SNR_hist_3G}.  
In total,  there are 63.0\%, 89.8\%, 97\%, 99.4\% \ac{BNS} mergers detectable to 3ET, 2ET1CE, 1ET2CE, 3CE respectively as listed in Table~\ref{loct_3G}.
For networks consisting of more \ac{CE}-like detectors such as 1ET2CE and 3CE, 
the distribution of \acp{SNR} peak at higher value,  which is largely due to the better sensitivity of \ac{CE} at higher frequencies. 

\subsubsection{Localization uncertainty}
\begin{figure*}[h]
\includegraphics[width=17cm]{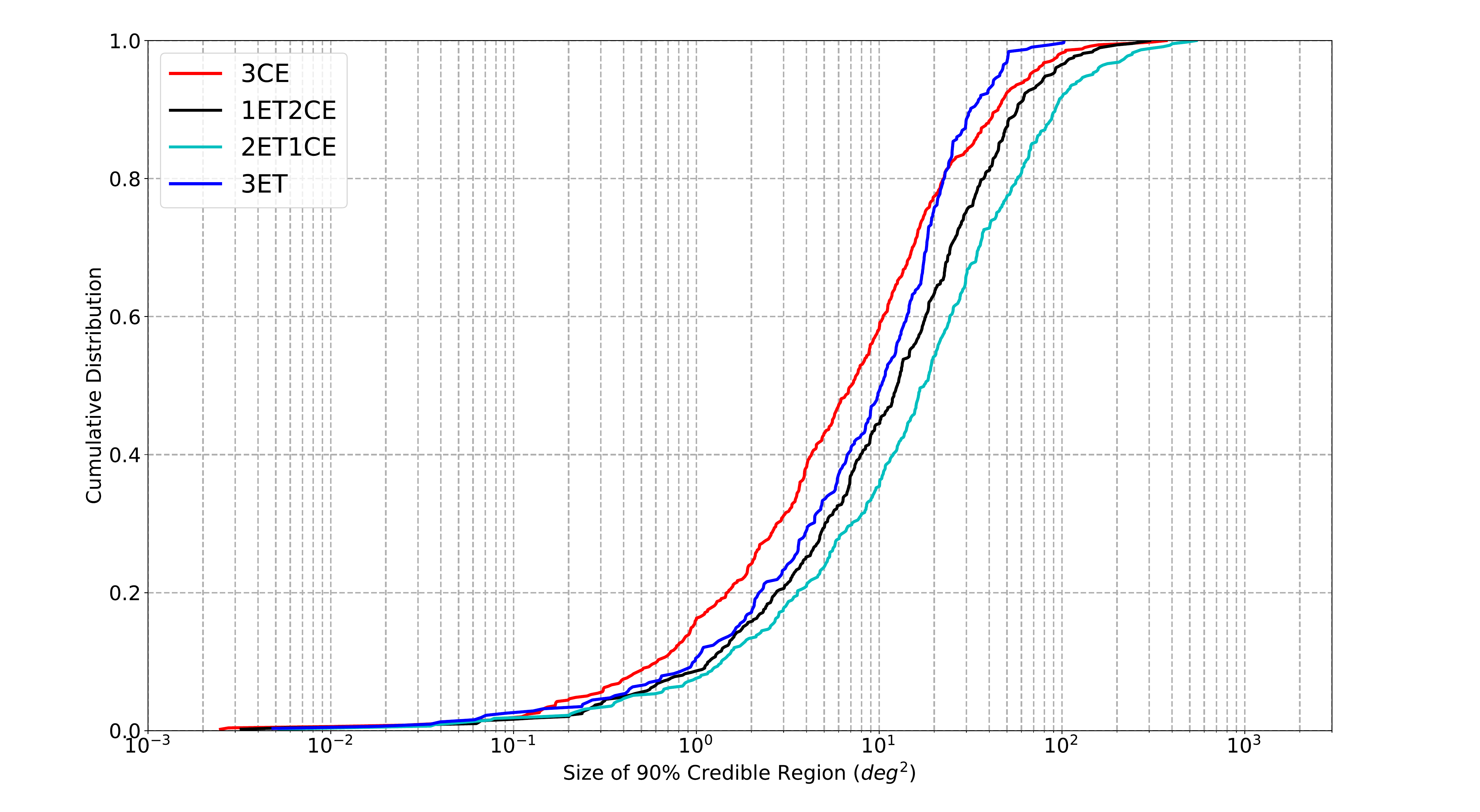}
\caption{The cumulative distribution of the size of the $90\%$ credible region for the \ac{BNS} mergers following the \ac{DTD}. The x-axis show the size of 90\% credible region in deg$^2$. }\label{ER_3G}

\subfigure[3ET] { \label{fig:a} 
\includegraphics[width=1\columnwidth]{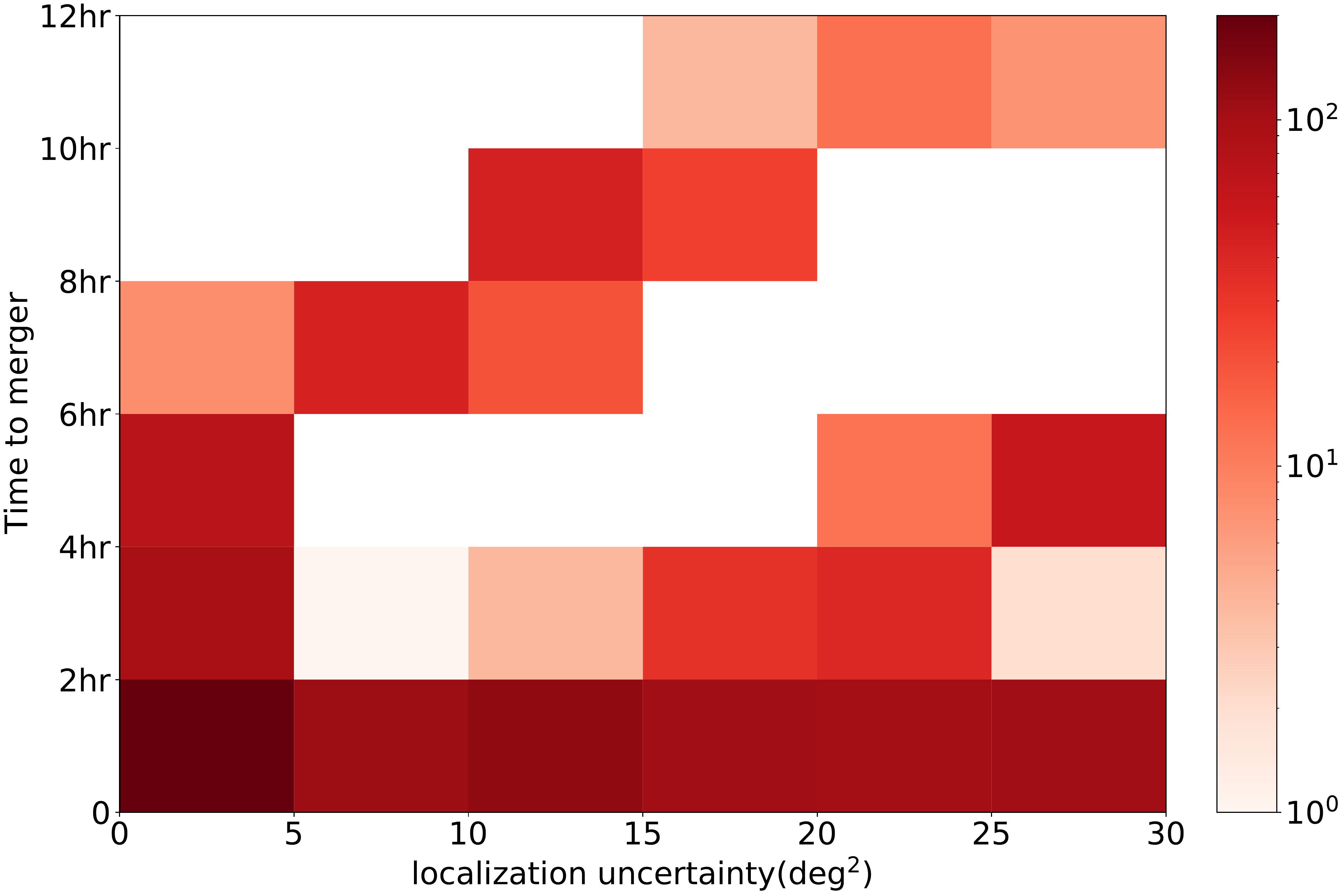}} 
\subfigure[2ET1CE] { \label{fig:b} 
\includegraphics[width=1\columnwidth]{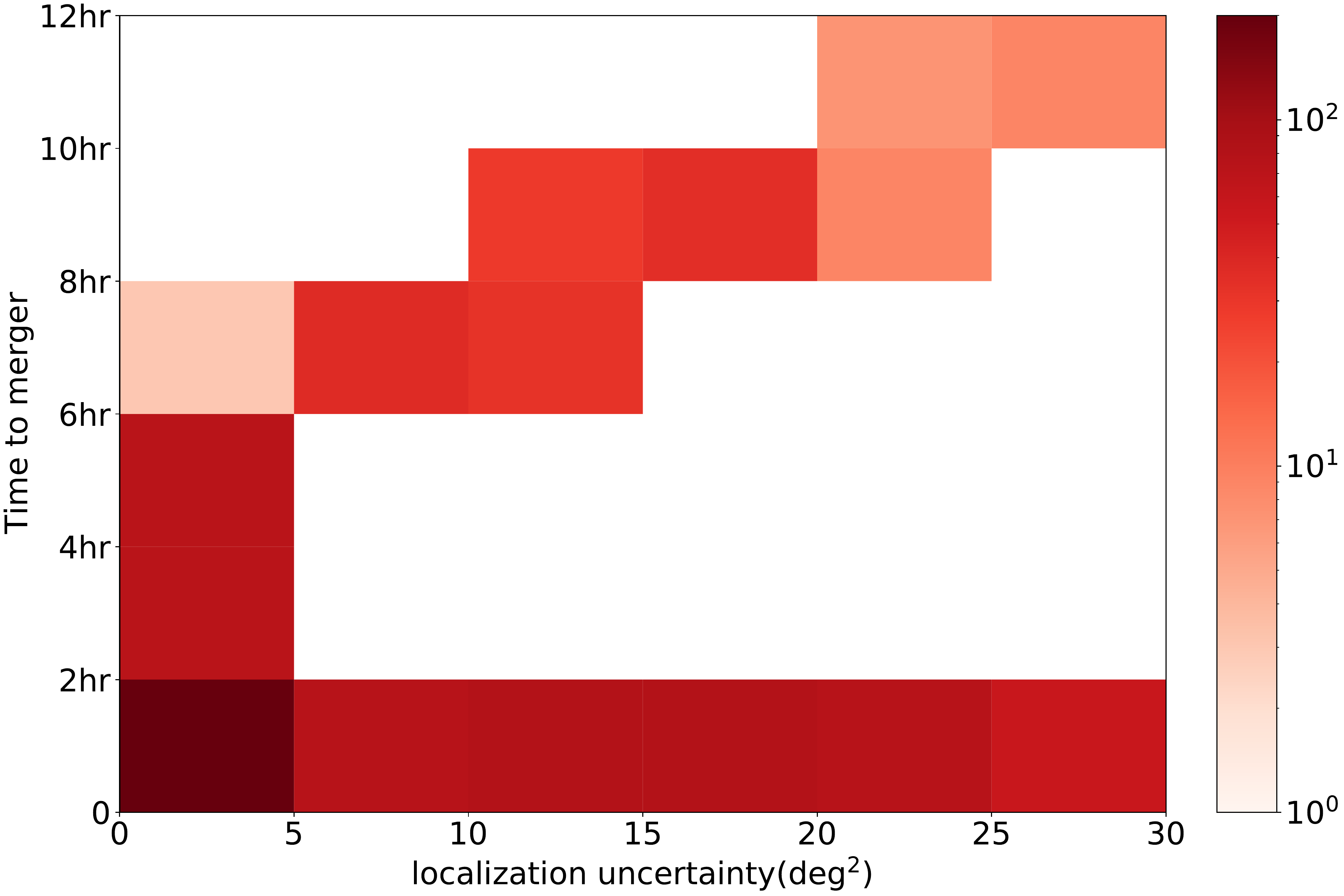}} 

\subfigure[1ET2CE] { \label{fig:c} 
\includegraphics[width=1\columnwidth]{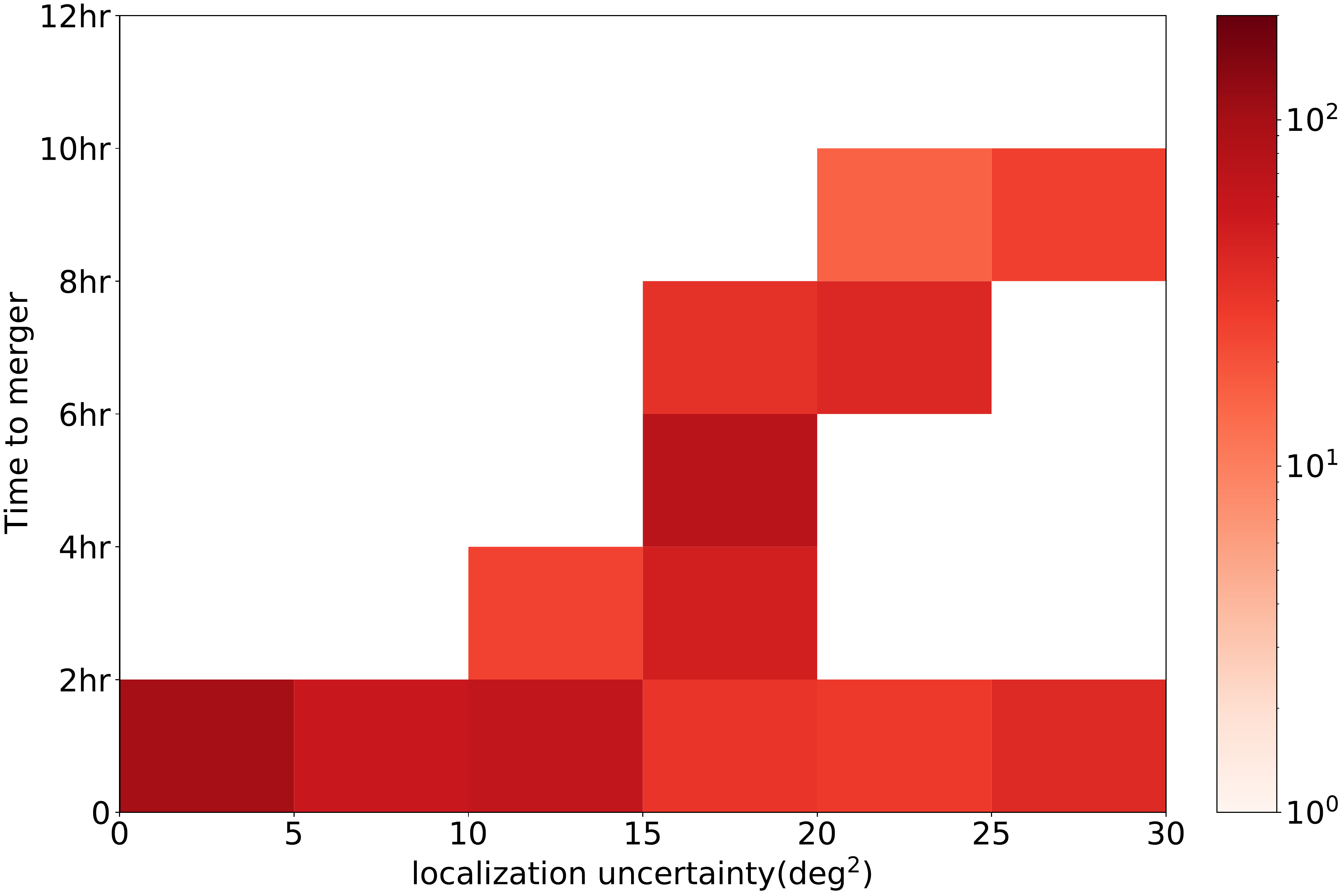}} 
\subfigure[3CE] { \label{fig:d} 
\includegraphics[width=1\columnwidth]{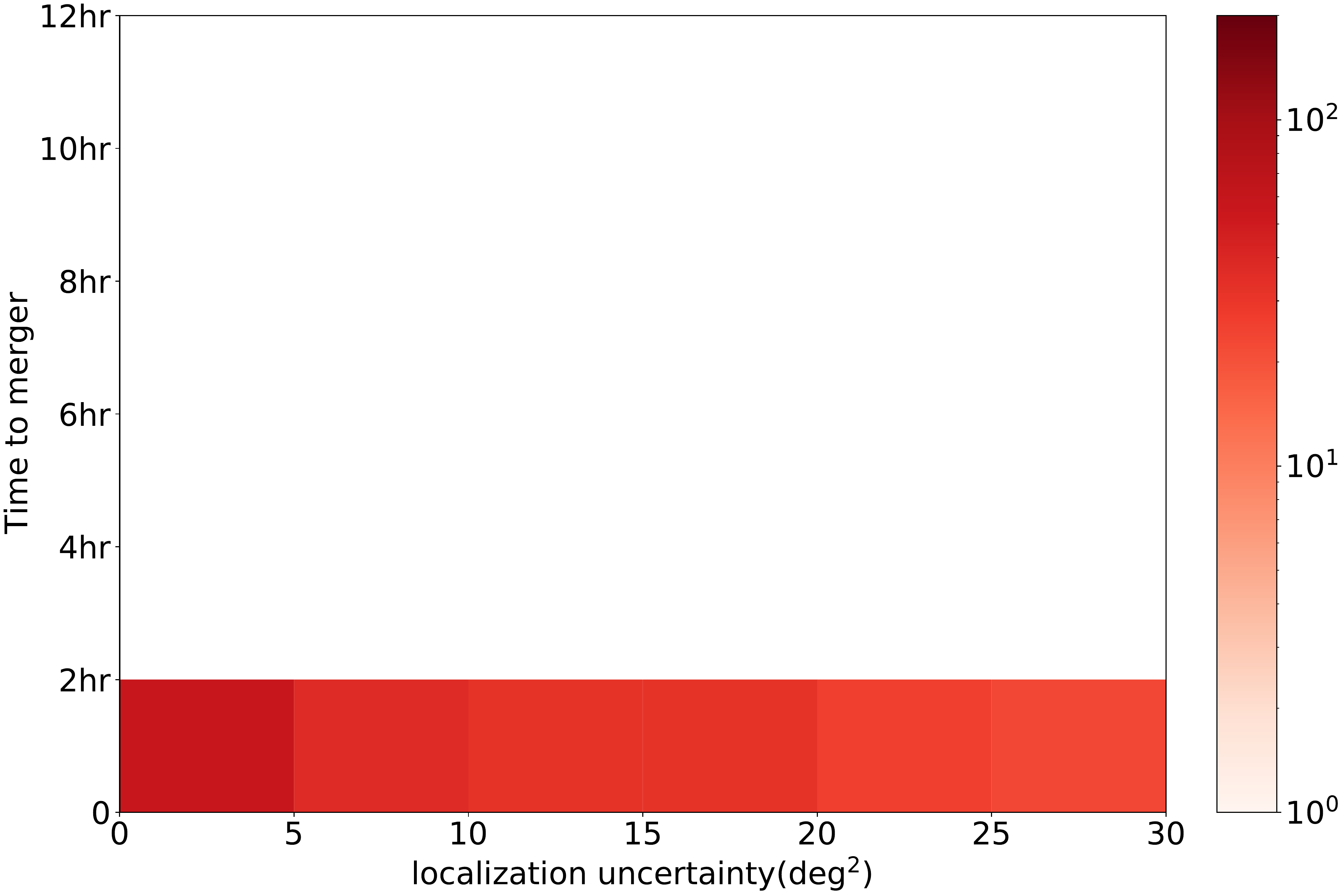}} 
\caption{Two-dimensional histograms showing the distributions of the time to merger
for \ac{BNS} mergers following \ac{DTD}. The horizontal axis of every subplot is the size of the $90\%$ credible region,  and the vertical axis represents the time to merger. 	
The color represents the number of sources which achieve 
the early warning criteria with the localisation requirement being $\leq 30\mathrm{deg^2}$ .}
\label{EW_3G_2d_30}
\end{figure*}

\indent For all simulated BNS sources following the \ac{DTD},  all detector networks can localize BNS signals within $\mathcal{O}\left(10^2\right)\text{deg}^2$.  
The cumulative distribution of size of the $90\%$ credible regions are shown in Figure~\ref{ER_3G}. 
It should be noted that only localization results of those detectable sources as described in Sec~\ref{snr_dtd} are presented in Figure~\ref{ER_3G}. As we can see, the curve of 2ET1CE, 1ET2CE, and 3CE have a trend of gradual change, toward a direction of occupying more proportion at smaller values of localization uncertainty. Behavior of 3ET is deviated from prediction of this gradual change, which originates from detectable source selection effect.
The upper limits of the localisation errors for the $90\%$, $50\%$ and $10\%$ best localised sources are shown in Table~\ref{loct_3G}. 
It could be found from Table~\ref{loct_3G} that for best localized 90\% detectable sources,  2ET1CE network could localize sources within $91.79\text{deg}^2$, and maximum localization uncertainty of 1ET2CE network shrinks by about 0.38 times compared to that of 2ET1CE network. In addition, 1ET2CE network could localize  
best localized half of detectable sources within $12.54\text{deg}^2$, which is about 0.28 times smaller than $17.57\text{deg}^2$ of 2ET1CE network. For the best localized 10\% of detectable sources, the differences are smaller, but the trend remains the same, 1ET2CE could localize sources within $1.17\text{deg}^2$ which is still less than $1.37\text{deg}^2$ of 2ET1CE network.
\begin{table}[]
  \begin{tabular}{p{1.34cm}p{1.34cm}p{1.34cm}p{1.34cm}p{1.34cm}}
      \hline
      \hline
        & 3ET ($\text{deg}^2$) & 2ET1CE ($\text{deg}^2$) & 1ET2CE ($\text{deg}^2$) & 3CE  ($\text{deg}^2$) \\
      \hline  
      &  63.0\% & 89.8\% & 97\% & 99.4\%\\            
       \hline
        90\% & $31.59$ & $91.79$ & $56.77$ & $44.68$\\
        50\%  & $10.25$& $17.57$ & $12.54$ &7.02 \\
        10\%  & $0.97$ & $1.37$ & $1.17$&0.61 \\     
      \hline
      \hline
  \end{tabular}
  \caption{A table showing the upper limits for the size of the 90\% credible regions for the best localized $90\%$, $50\%$ and $10\%$ of the detectable sources following \ac{DTD}. The second row denotes the percentage of the detectable sources for the detector networks. \\}
  \label{loct_3G}
  \end{table} 
  
\subsubsection{Early warning capabilities}
The results are presented in Figure~\ref{EW_3G_2d_30} as a two-dimensional histogram showing the distribution of time to merger versus the size of $90\%$ credible region for \ac{BNS} mergers following DTD. The chosen value of $30\text{deg}^2$ is motivated by the fact that all simulated events at $1600$Mpc can be localised to within such an area with all simulated networks.
For comparison, we also employ $100\text{deg}^2$ as the value for the second condition. The result is shown in Appendix.

From Figure~\ref{EW_3G_2d_30}, 3ET network and 2ET1CE network both could detect several BNS signals more than 10 hours before merger with localization uncertainty smaller than $\mathrm{30deg^2}$. 
In terms of the detectable source quantity, more CE included network could detect more sources which achieve the early warning criteria. 
But in terms of the time to merger distribution, it is obvious that early warning of more ET included network are more widely distributed and peaked at higher value in vertical direction. For example, for 1ET2CE network, which replaces one CE with ET in 3CE network,  the upper right corner of the figure fills in the blanks, which means joining of ET improves the early warning performance in some extent, and could give an early warning alert as early as 10 hours. Besides, compared to 1ET2CE network, 2ET1CE network could still detect some signals as early as 10 hours while with smaller localization uncertainty, which indicates a greater opportunity to perform successful \ac{EM} follow-up observations.\\
\section{Conclusion and Discussion}\label{final}
GW detections are rapidly evolving as a routine tool in the multi-messenger era. Maximizing GW's scientific output by associating with its EM counterpart is our major concern in multi-messenger astronomy. Localization and early warning performance of GW detector largely affects the efficiency of EM follow-up operation. In this paper, we calculated and compared the localization uncertainty and early warning performance of four third generation ground based GW detector networks for BNS sources at fixed distance and BNS sources following \ac{DTD}.  The duration of simulated signals is about 5 days long, therefore we included the earth rotation effect. And considering that during detection, the detector has motion relative to source, thus we also included this Doppler effect. Besides \ac{ET} to be built in Europe, and \ac{CE} to be built in North America, we also simulated four detector networks involving combinations of the aforementioned detectors and the following: ET in Australia (ET-A), CE in Australia (CE-A), ET in North America and CE in Italy.  The comparison of localization capabilities of ET-A included network and CE-A included network is our major concern.

In terms of SNR, for a \ac{BNS} population following the DTD approach, networks with more \ac{CE} detectors tends to have higher SNRs due to \ac{CE} having better sensitivity at medium to high frequency band, lead to a smaller source localization uncertainty. Using Fisher matrix analysis, 90\% of the best localized sources can be localized within an area of respectively $31.59\text{deg}^2$, $91.79\text{deg}^2$, $56.77\text{deg}^2$ and $44.68\text{deg}^2$ for 3ET, 2ET1CE, 1ET2CE, 3CE network. The 3ET network seems to give an excellent source localization estimate but only because poorly localized sources are rejected due to their SNR not being above the required threshold of 8. For \ac{BNS} mergers at fixed distances, the comparison of localization performance between tested networks remain basically similar to that of \ac{BNS} mergers following \ac{DTD}.

In terms of time to merger distribution, for \ac{BNS}s at 200 Mpc, all tested networks achieved the early warning criteria of $30\text{deg}^2$ at least 100 seconds before merger, and the time to merger distribution is widely distributed in range of 100 seconds to 10 hours, which indicates a bright future for successful \ac{EM} follow-up observations in third generation \ac{GW} detector era. And for both \ac{BNS} mergers at fixed distances and following \ac{DTD}, more ET included network is more likely to detect BNS signals earlier because the better sensitivity of ET in low frequency band, leading to \ac{BNS} signals being in-band for longer. 

Overall, compared with the 2ET1CE network, the 1ET2CE network detects more BNS signals with achieving a better localization uncertainty. However, the 2ET1CE network tends to giving alert earlier than the 1ET2CE network. All things considered, we think building a CE-like detector in Australia to create the 1ET2CE network is a reasonable compromise, as this would lead to an excellent localization performance without losing too much early warning performance compared to 2ET1CE network. 
\\
\acknowledgments
We are grateful to Prof. Mohammadtaher Safarzadeh and Prof. Seyed Alireza Mortazavi for helpful discussion about delay time distribution of binary neutron stars. We would like to thank Jan Harms and Christopher Berry for reviewing the paper during the internal LIGO review. We are also grateful for computational resources provided by Cardiff University, and funded by an STFC grant supporting UK Involvement in the Operation of Advanced LIGO. Y. L acknowledges scholarship provided by the University of Chinese Academy of Sciences (UCAS).  Y. L is supported by the National Program on Key Research and Development Project through grant No. 2016YFA0400804, and by the National Natural Science Foundation of China with grant No. Y913041V01, and by the Strategic Priority Research Program of the Chinese Academy of Sciences through grant No. XDB23040100.

\begin{appendices}
      \section{Appendix}
\begin{figure*}[h]
\centering
\includegraphics[width=15cm]{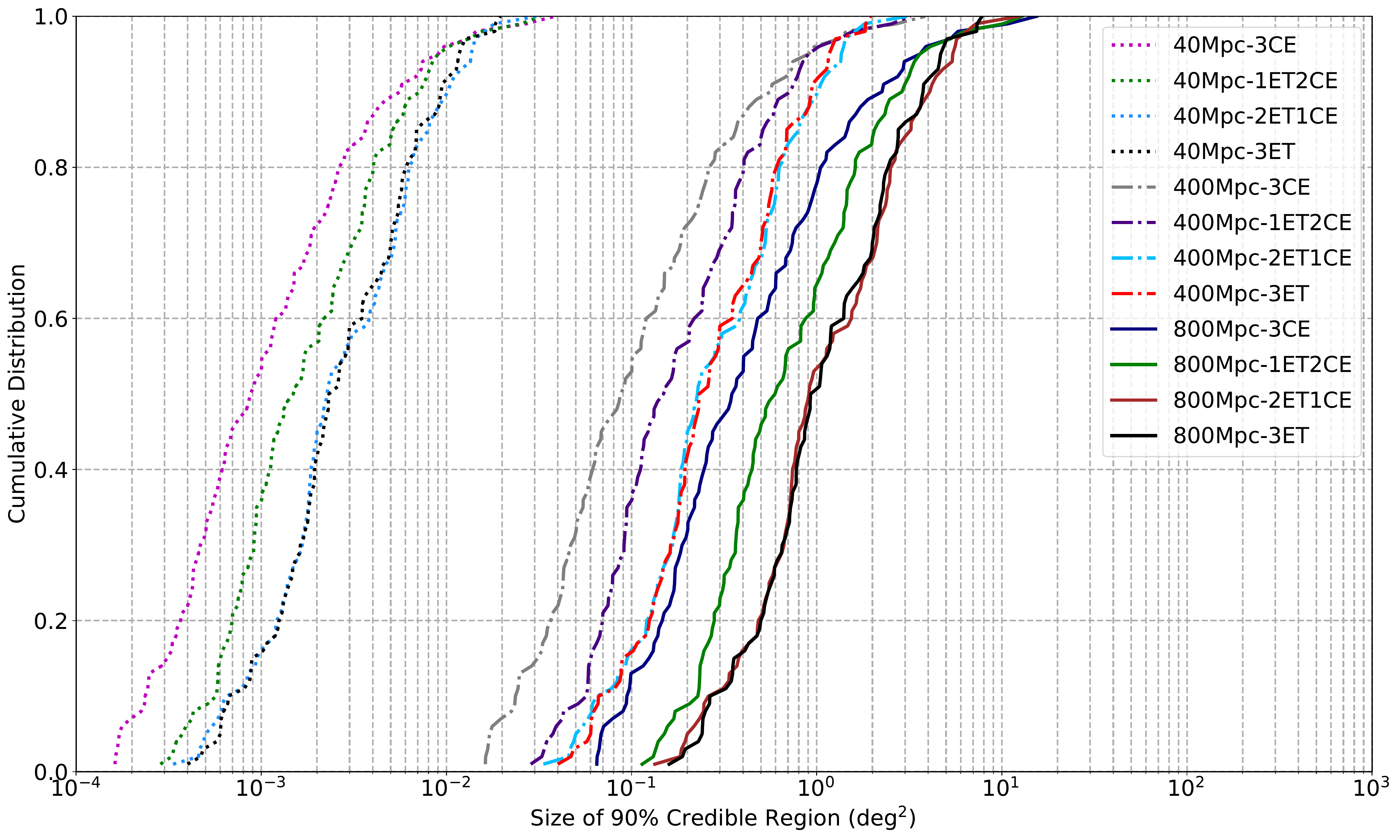}
\caption{The cumulative distributions of the size of the $90\%$ credible regions for the \ac{BNS} mergers at 40 Mpc, 400 Mpc and at 800 Mpc.  The x-axes show the size of the 90\% credible region and the upper limit of the x-axes corresponds to the size of the whole sky. }\label{er_appendix}
\end{figure*}

\begin{figure*}[h]
\subfigure[$1\text{deg}^2$] { \label{fig:a} 
\includegraphics[width=0.5\columnwidth]{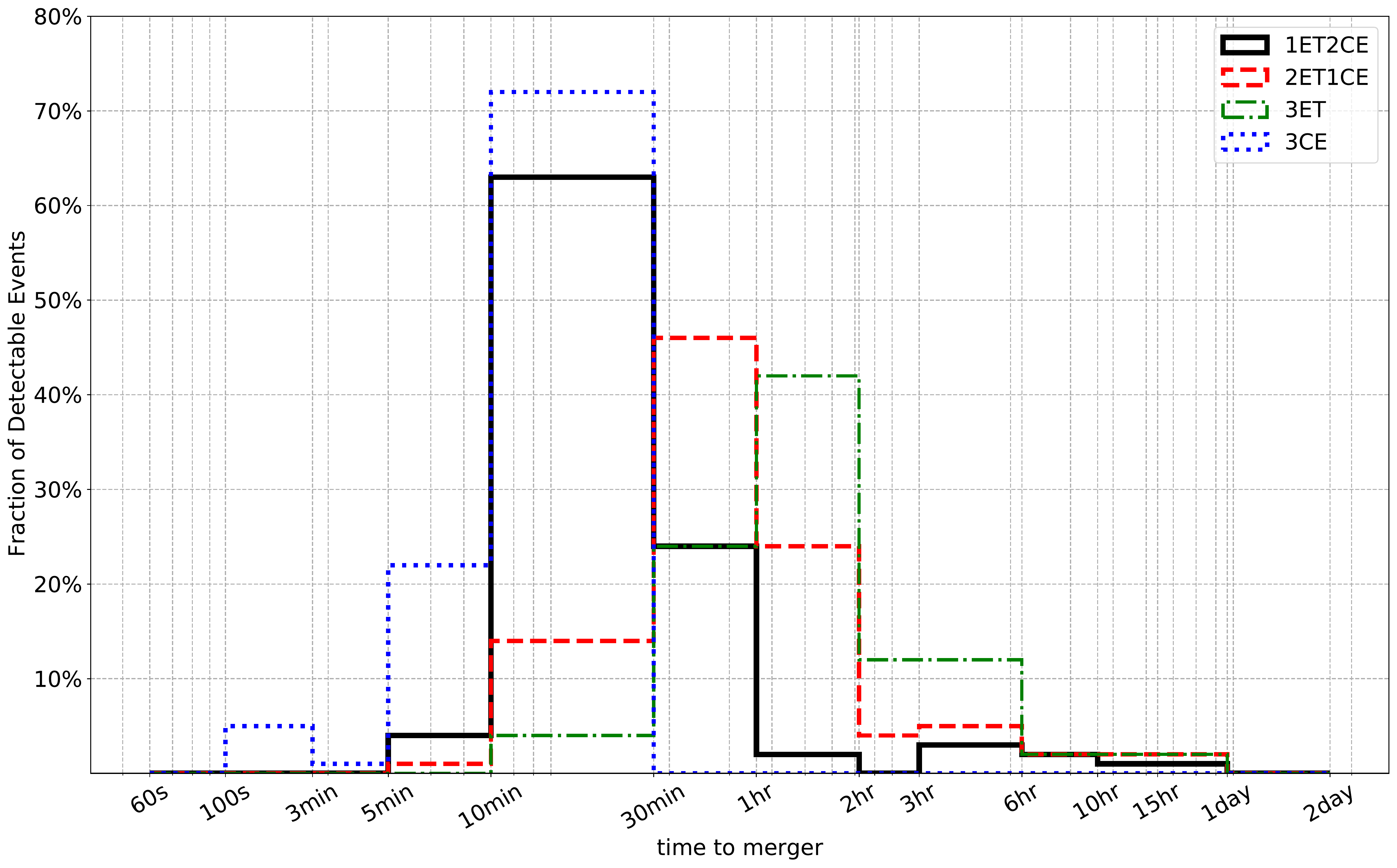}} 
\subfigure[$5\text{deg}^2$] { \label{fig:b} 
\includegraphics[width=0.5\columnwidth]{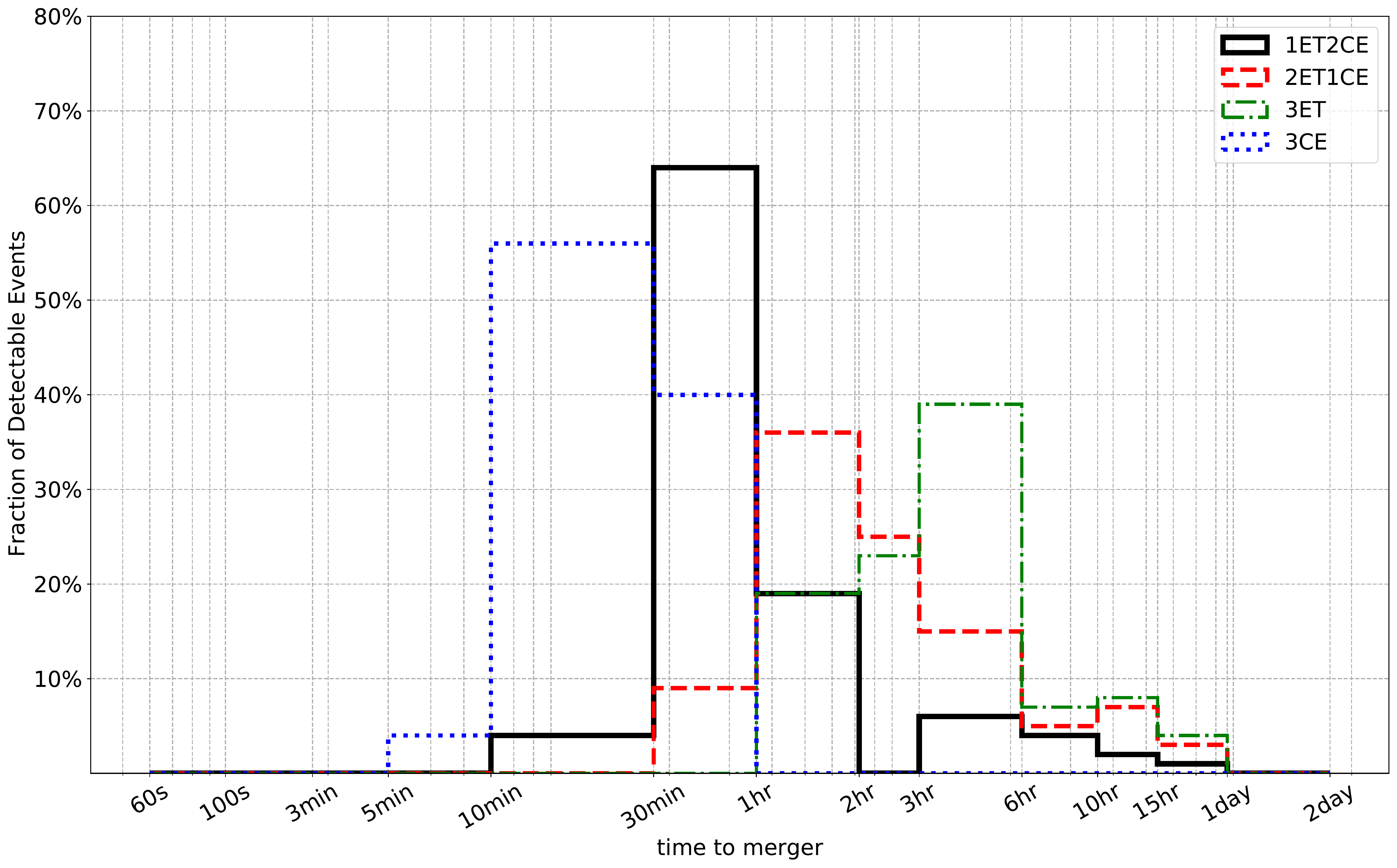}} 
\subfigure[$10\text{deg}^2$] { \label{fig:c} 
\includegraphics[width=0.5\columnwidth]{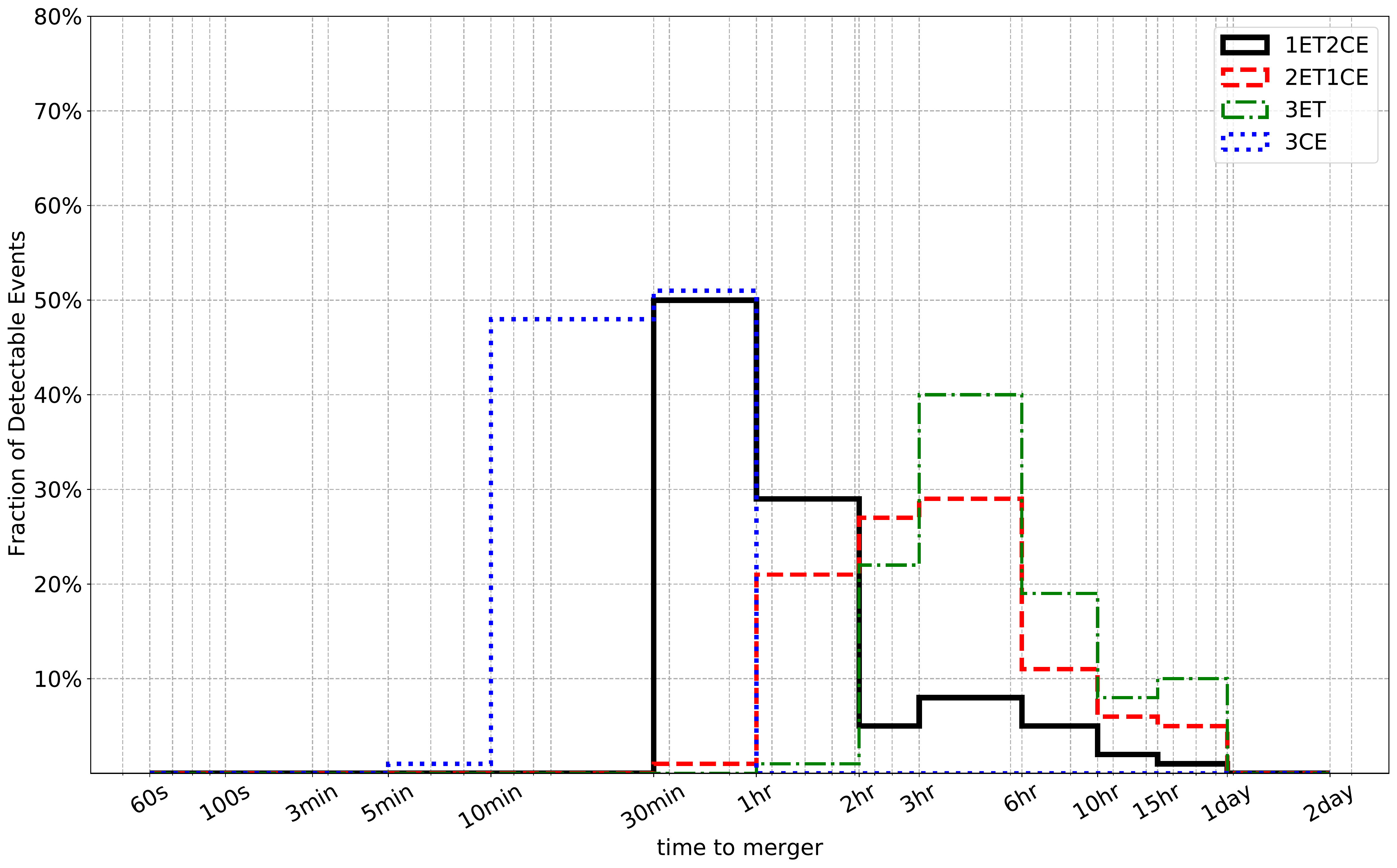}} 
\subfigure[$30\text{deg}^2$] { \label{fig:d} 
\includegraphics[width=0.5\columnwidth]{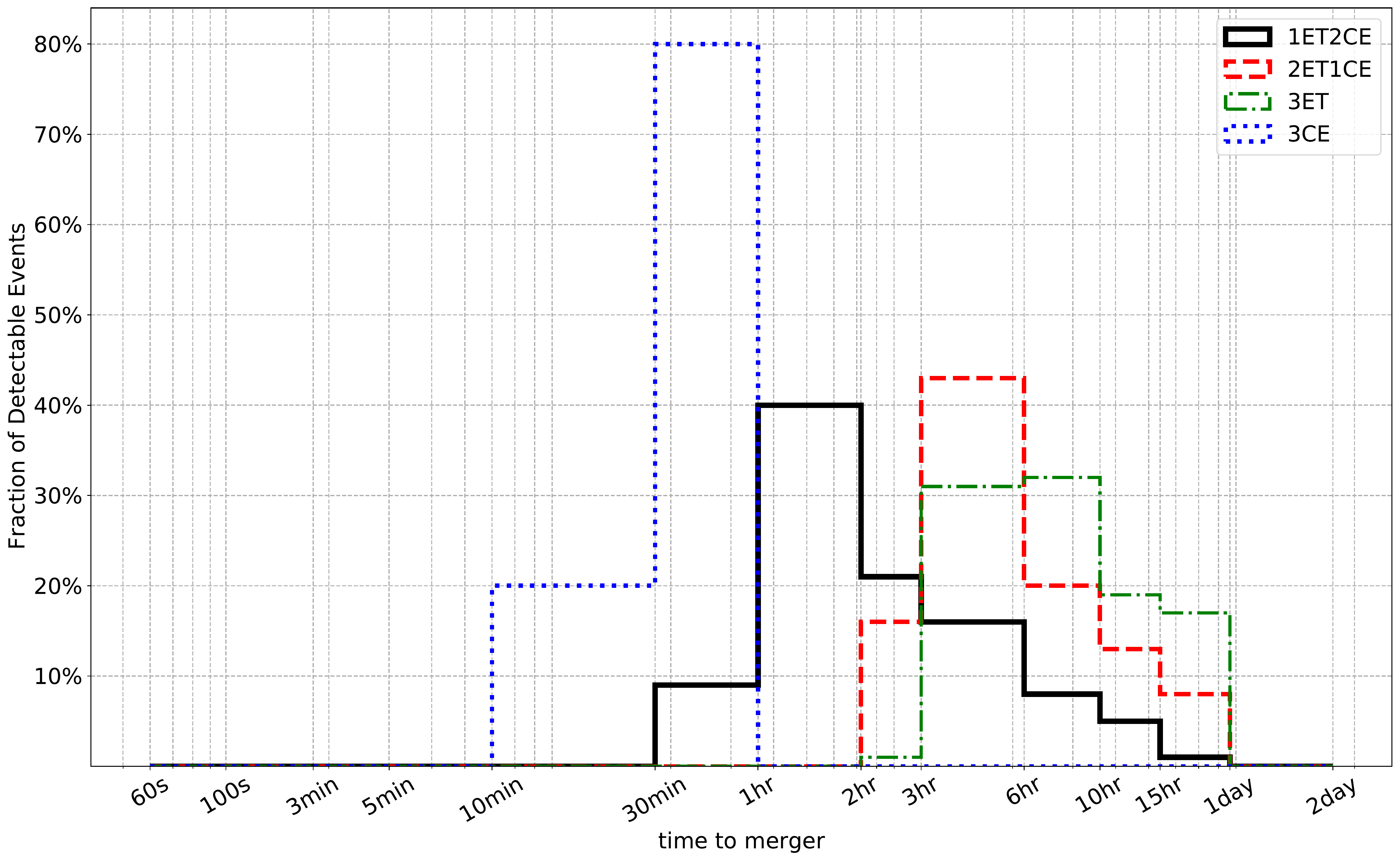}} 
\caption{The histogram distribution of time to merger by 4 detector networks for BNS at  40 Mpc with required localization uncertainty denoted as caption of each subplot. }\label{40EW}
\end{figure*}

\begin{figure*}[h]
\subfigure[$1\text{deg}^2$] { \label{fig:a} 
\includegraphics[width=0.5\columnwidth]{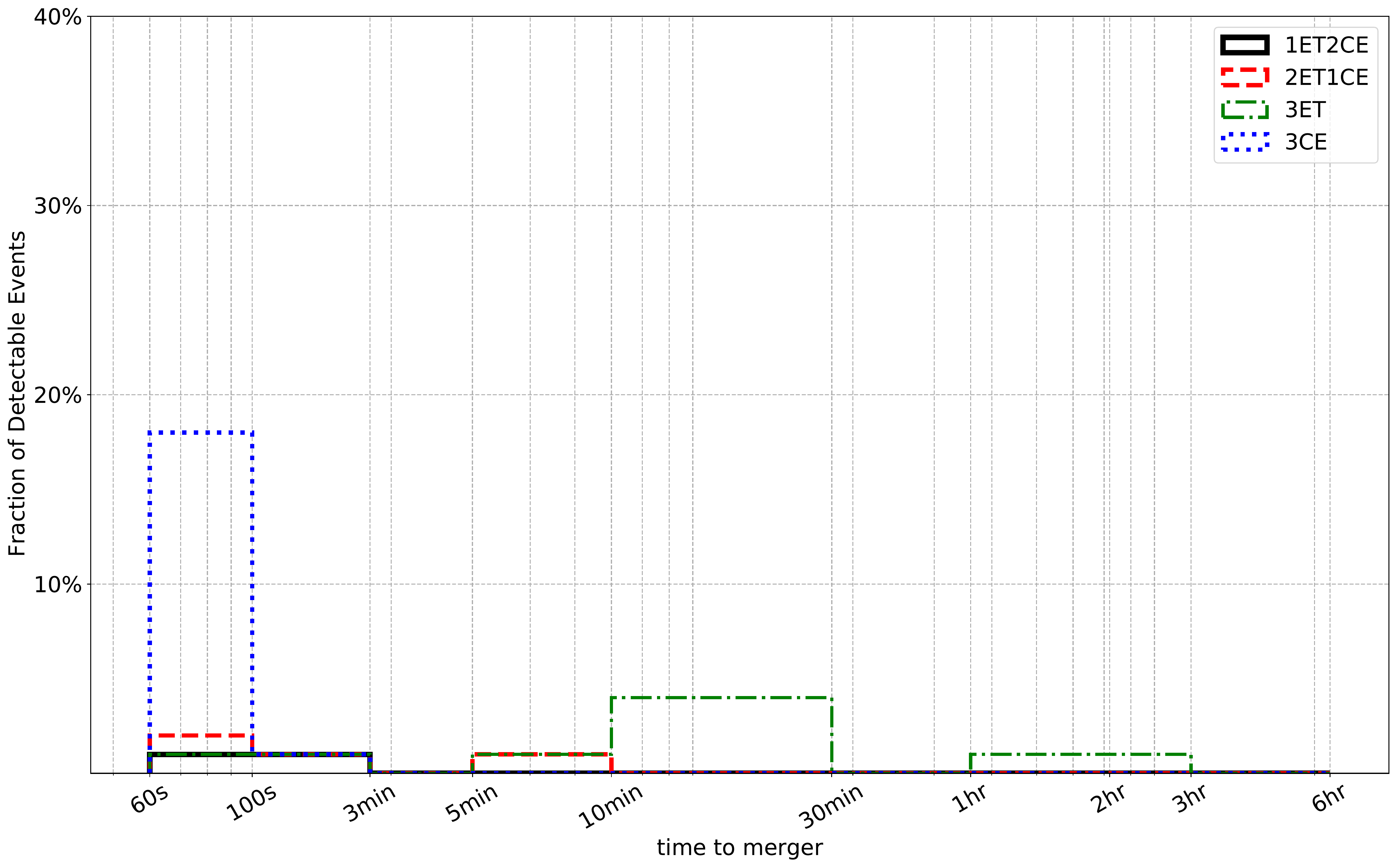}} 
\subfigure[$5\text{deg}^2$] { \label{fig:b} 
\includegraphics[width=0.5\columnwidth]{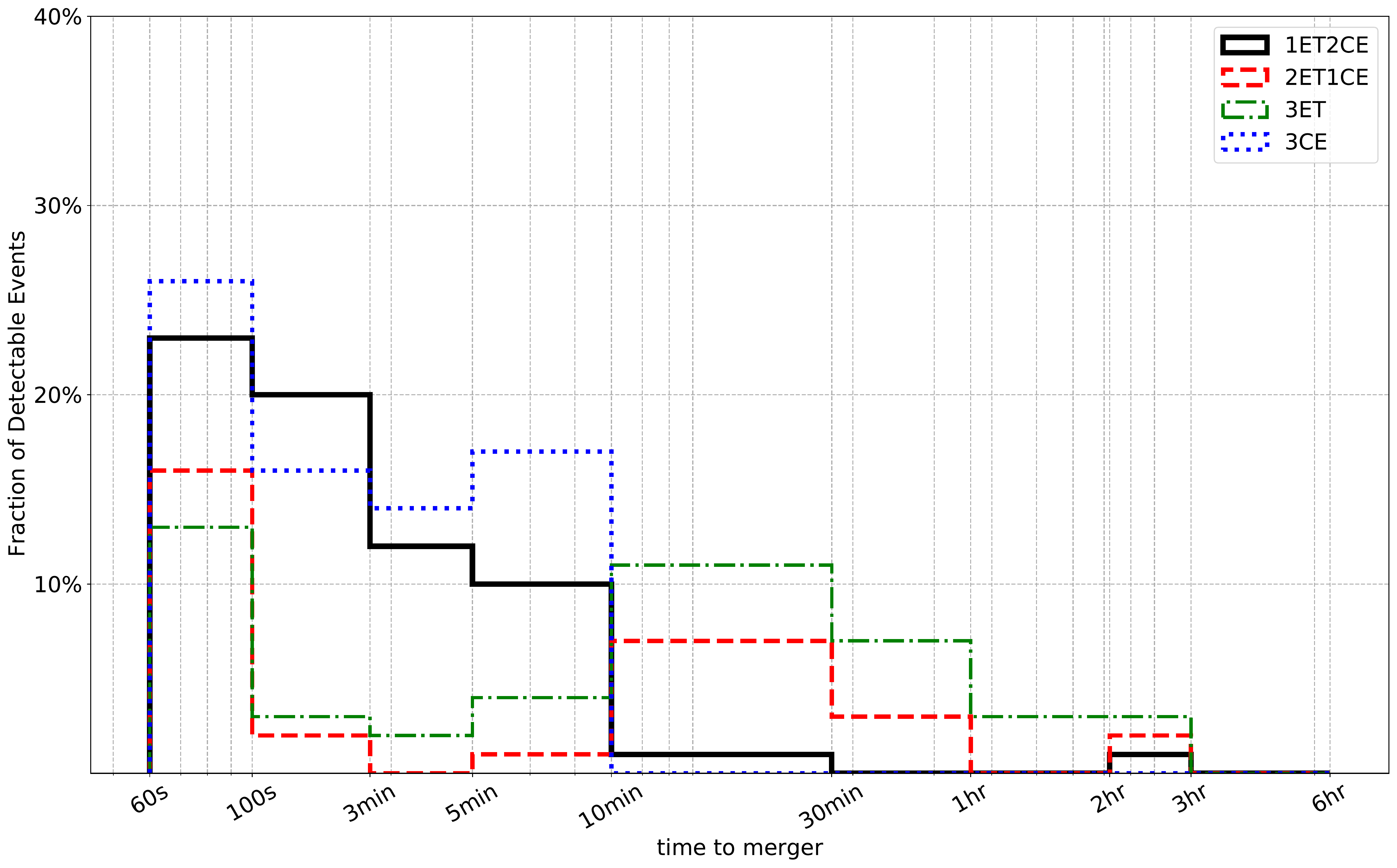}} 
\subfigure[$10\text{deg}^2$] { \label{fig:c} 
\includegraphics[width=0.5\columnwidth]{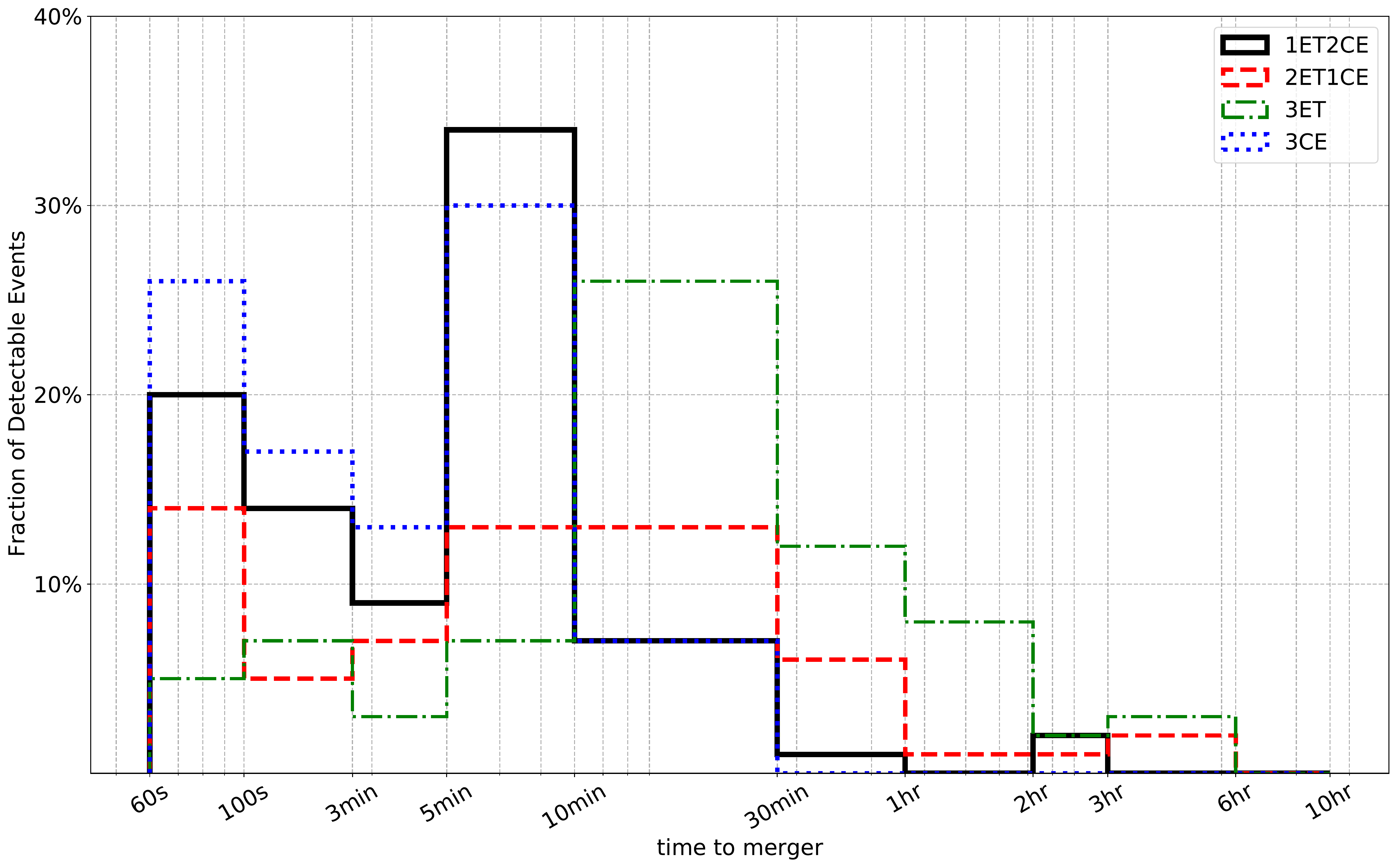}} 
\subfigure[$30\text{deg}^2$] { \label{fig:d} 
\includegraphics[width=0.5\columnwidth]{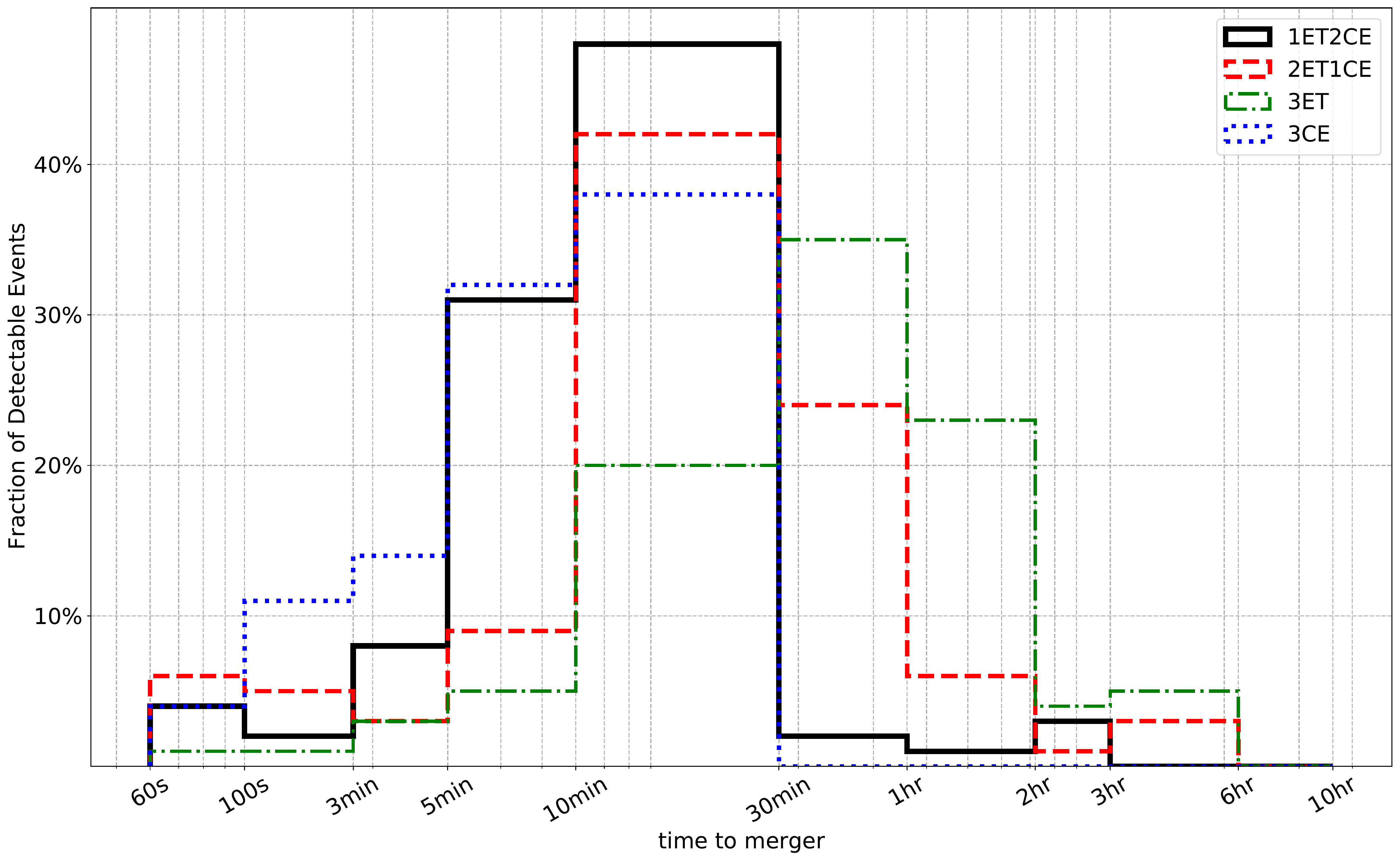}} 
\caption{The histogram distribution of time to merger by 4 detector networks for BNS at  400 Mpc with required localization uncertainty denoted as caption of each subplot.  }\label{400EW}
\end{figure*}

\begin{figure}[h]
\subfigure[$5\text{deg}^2$] { \label{fig:a} 
\includegraphics[width=0.5\columnwidth]{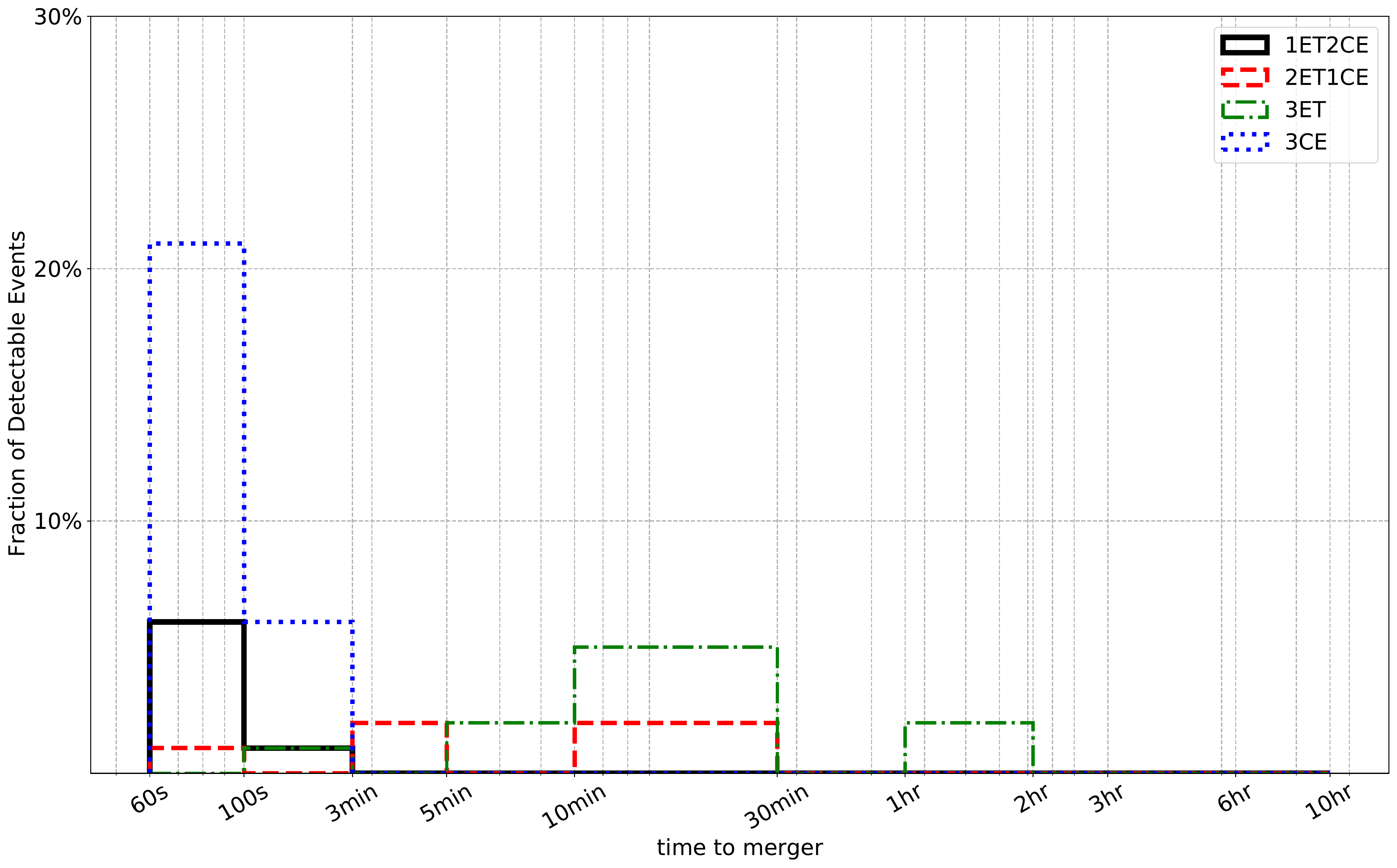}} 
\subfigure[$10\text{deg}^2$] { \label{fig:a} 
\includegraphics[width=0.5\columnwidth]{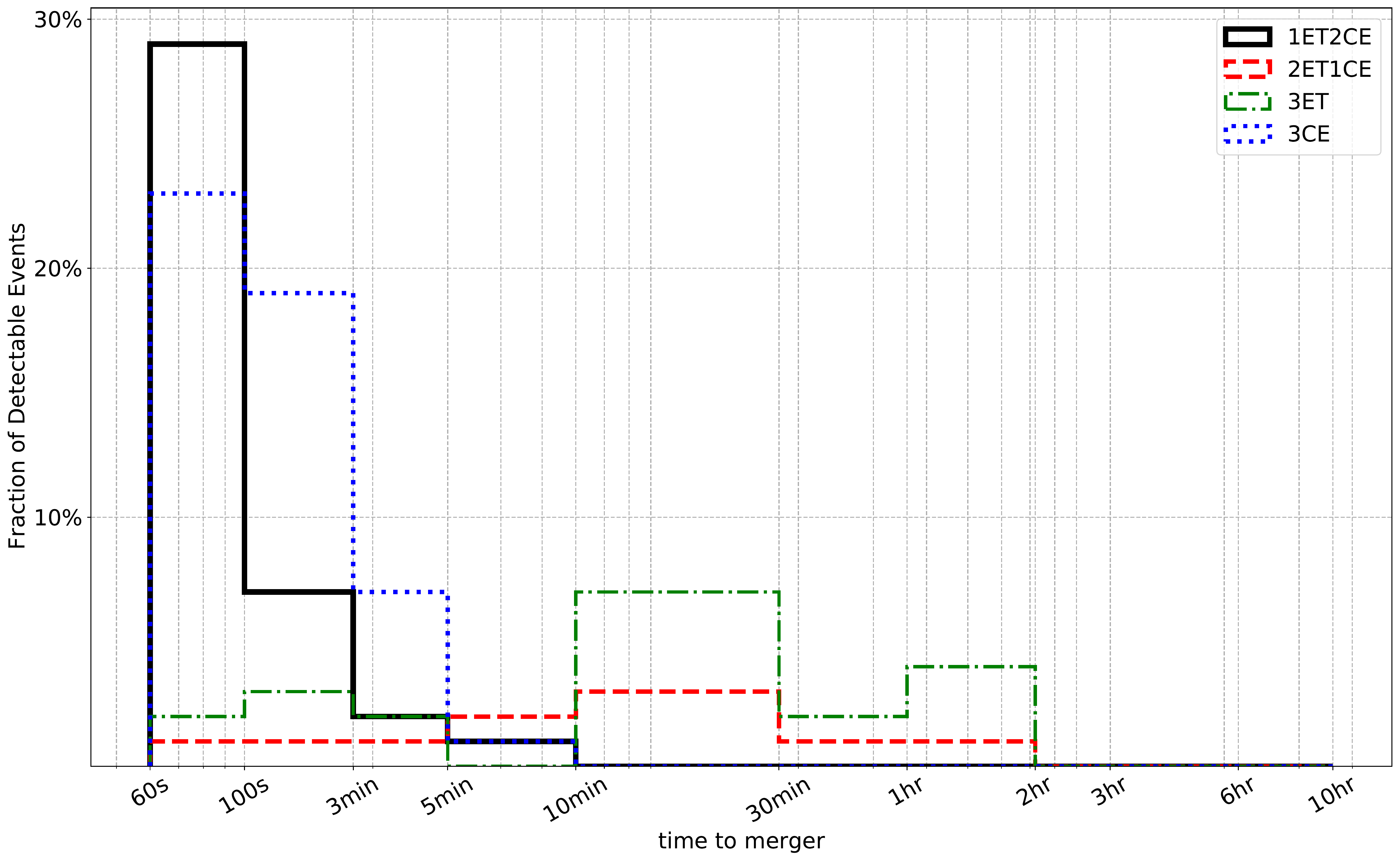}} 
\subfigure[$30\text{deg}^2$] { \label{fig:b} 
\includegraphics[width=0.5\columnwidth]{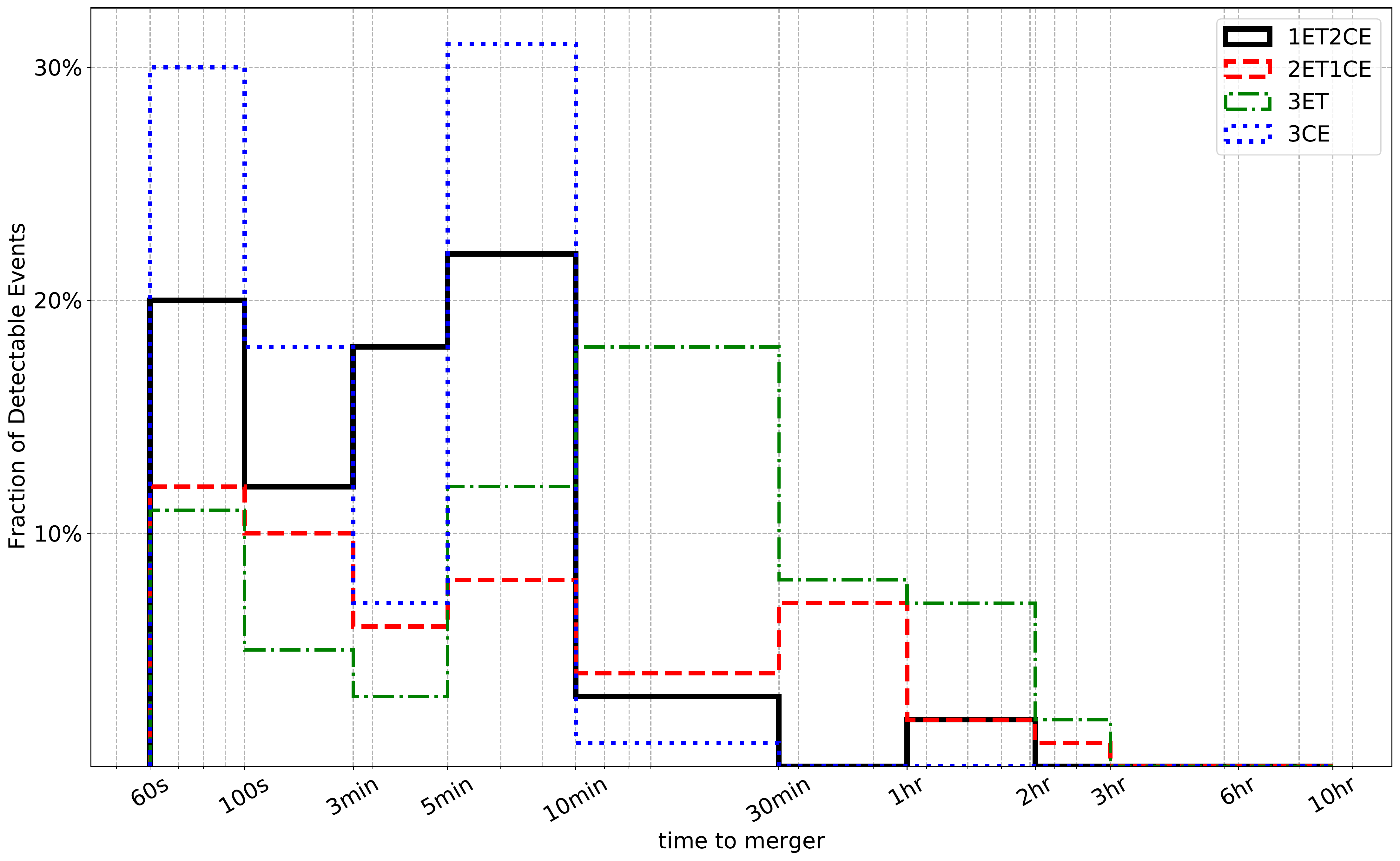}} 
\caption{The histogram distribution of time to merger by 4 detector networks for BNS at  800 Mpc with required localization uncertainty denoted as caption of each subplot. }\label{800EW}
\end{figure}

\begin{figure}[h]
\subfigure[$10\text{deg}^2$] { \label{fig:a} 
\includegraphics[width=0.5\columnwidth]{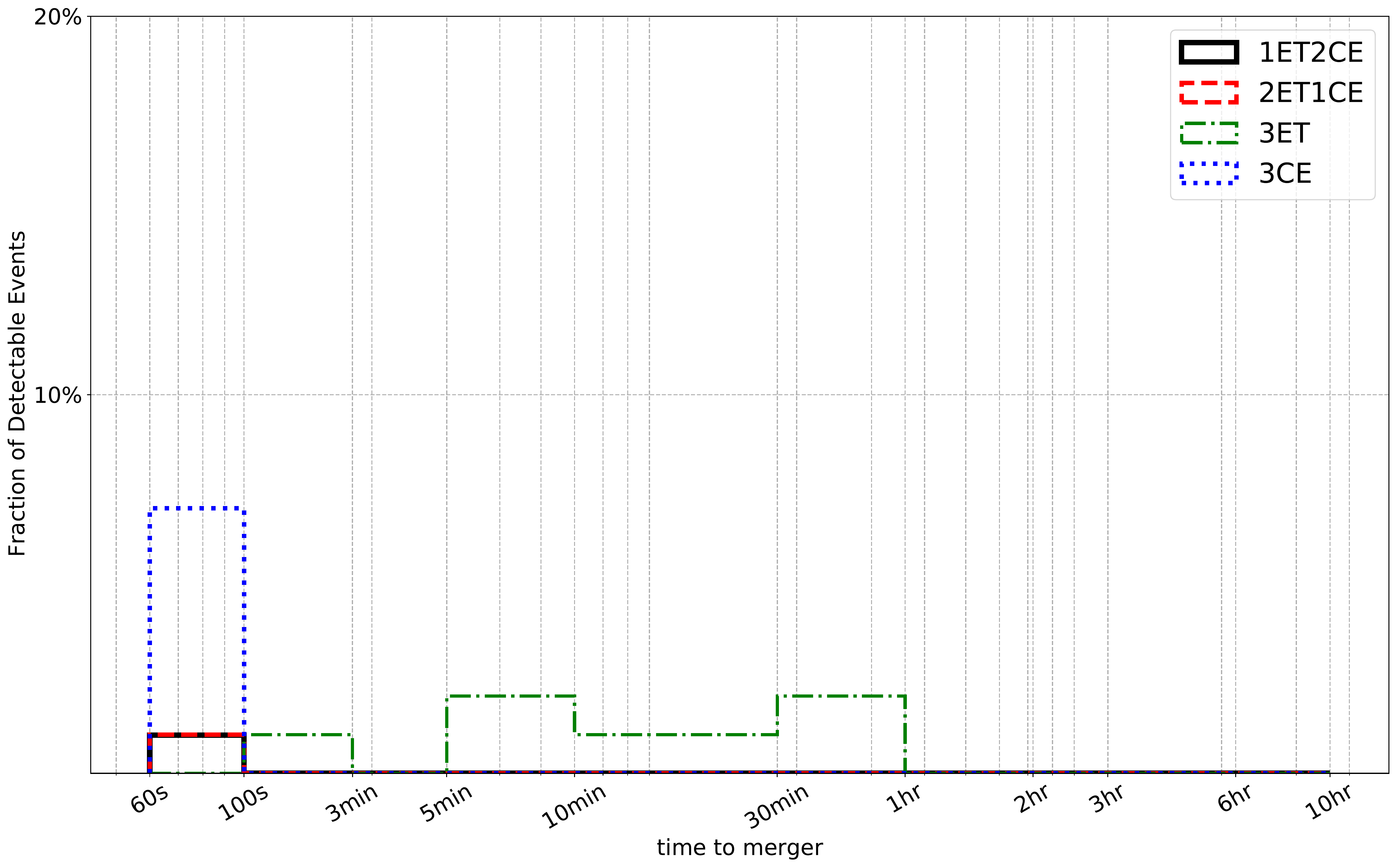}} 
\subfigure[$30\text{deg}^2$] { \label{fig:b} 
\includegraphics[width=0.5\columnwidth]{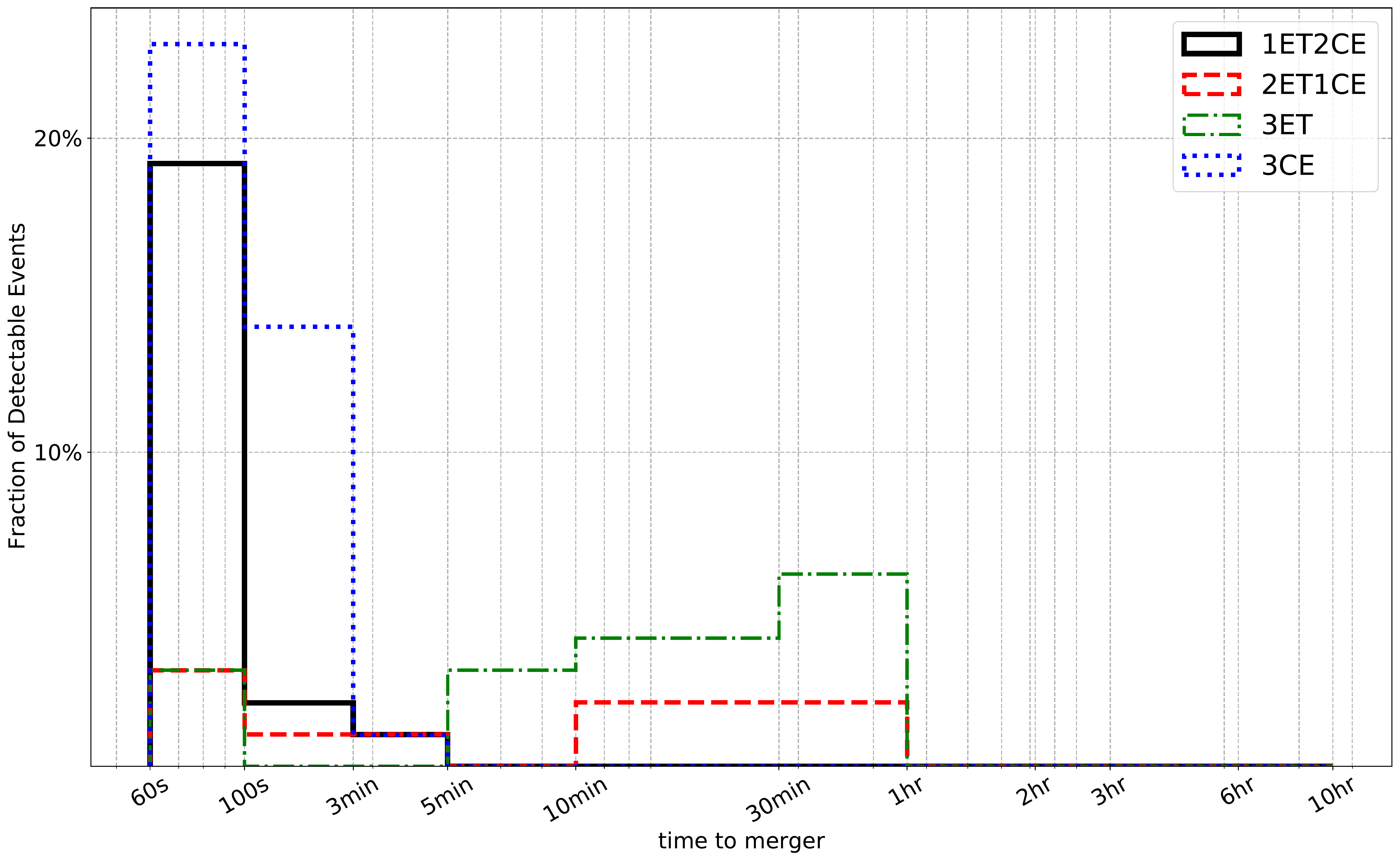}} 
\caption{The histogram distribution of time to merger by 4 detector networks for BNS at  1600 Mpc with required localization uncertainty denoted as caption of each subplot. }\label{1600EW}
\end{figure}

\begin{figure}[h]
\subfigure[3ET] { \label{fig:a} 
\includegraphics[width=0.5\columnwidth]{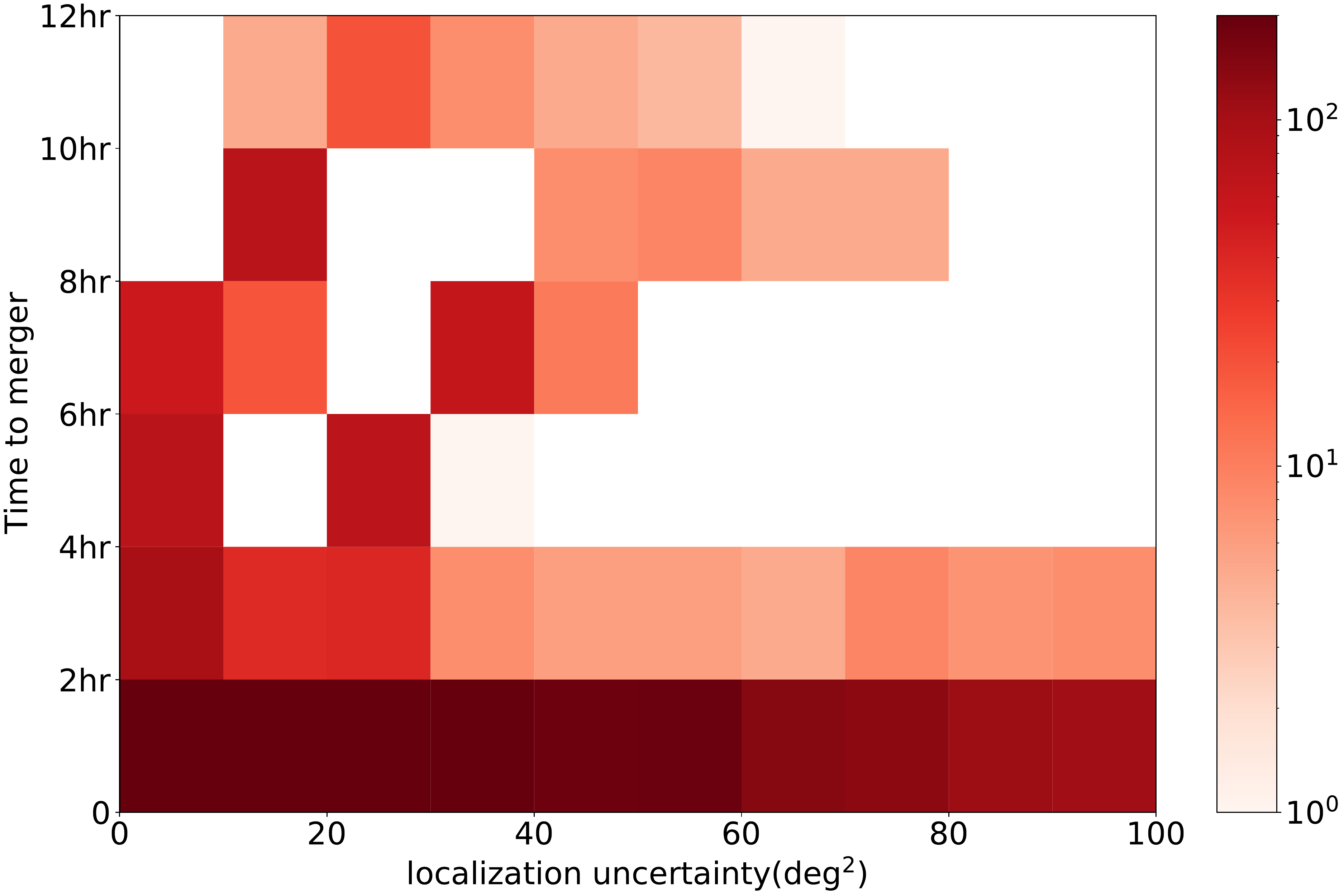}} 
\subfigure[2ET1CE] { \label{fig:b} 
\includegraphics[width=0.5\columnwidth]{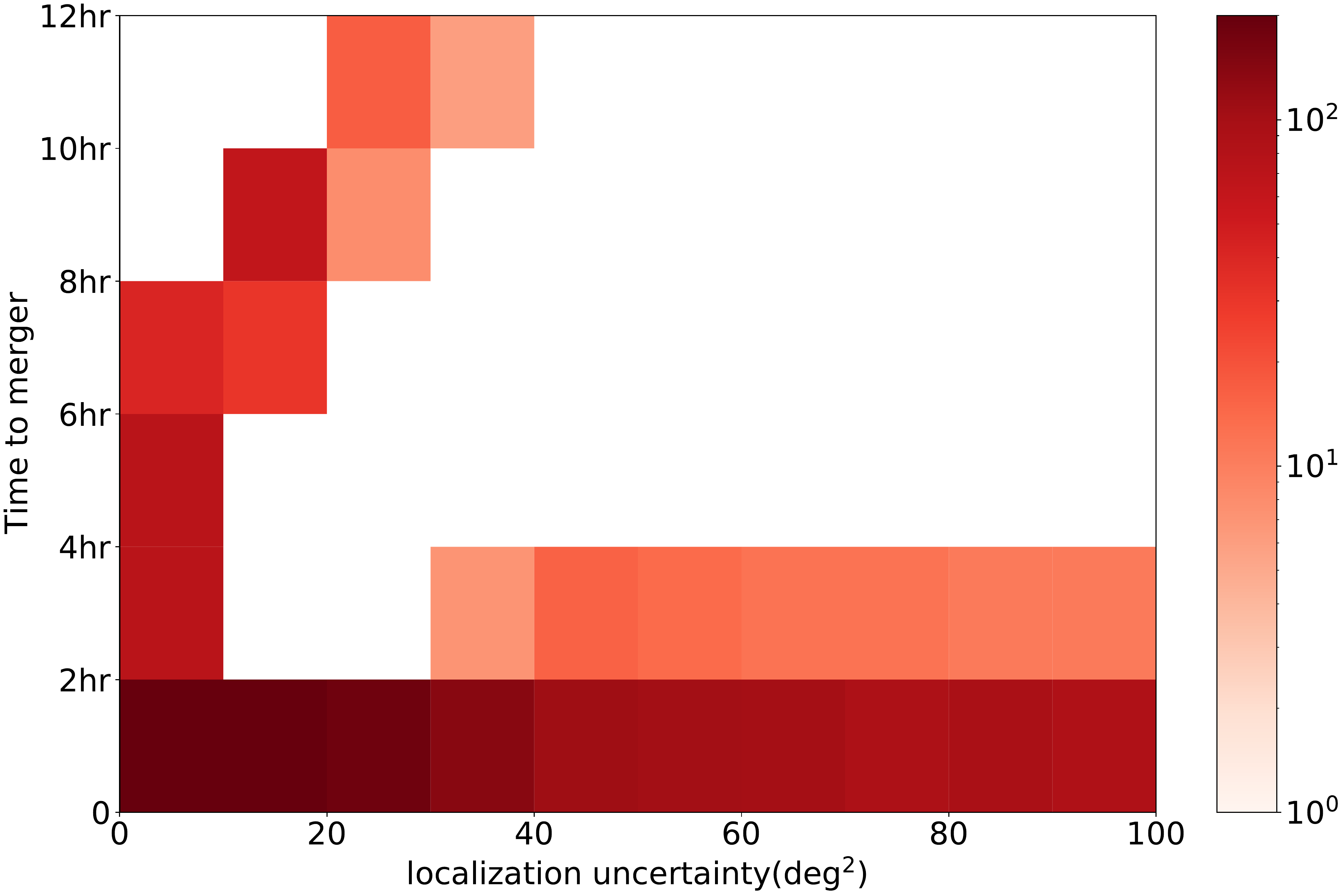}} 

\subfigure[1ET2CE] { \label{fig:c} 
\includegraphics[width=0.5\columnwidth]{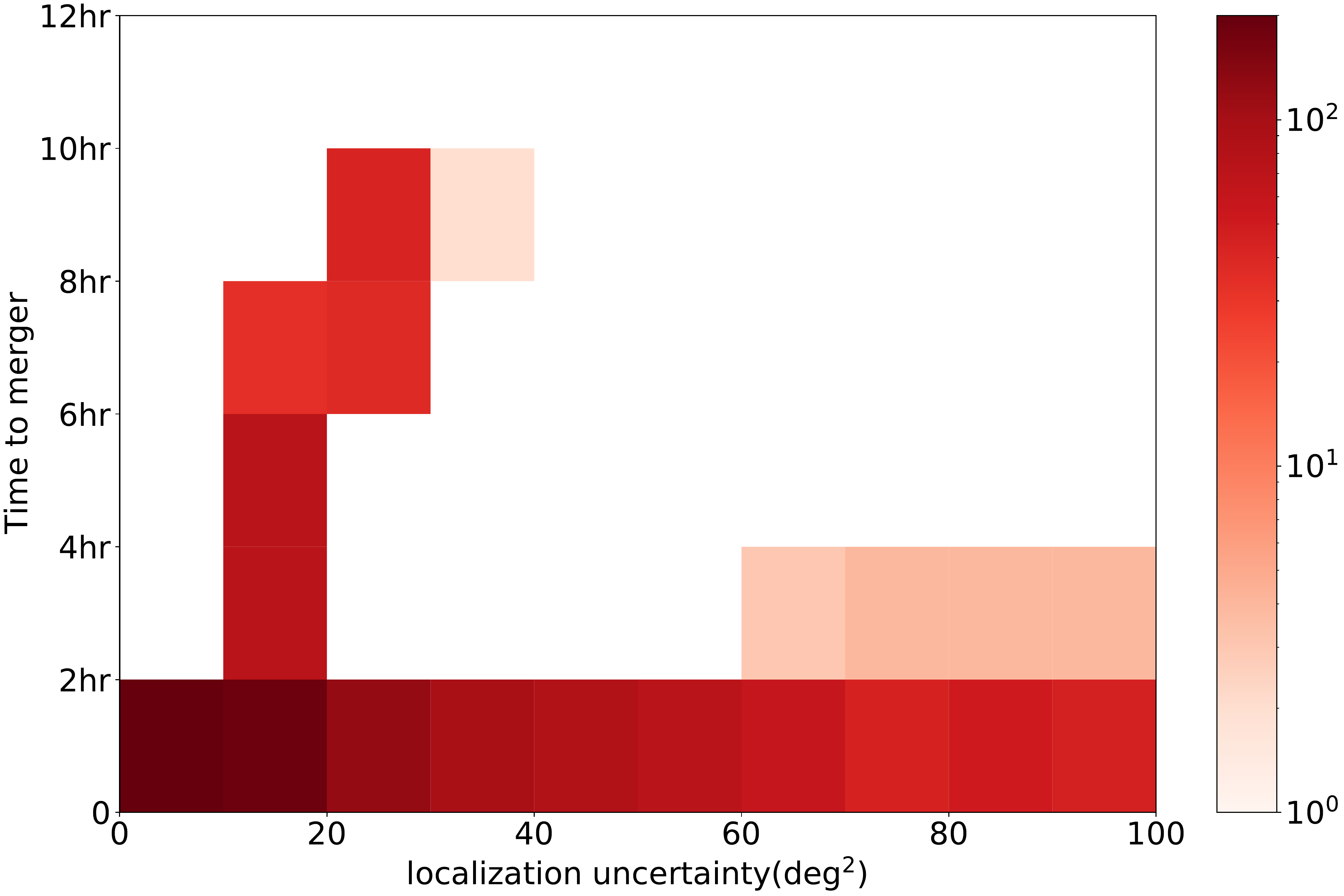}} 
\subfigure[3CE] { \label{fig:d} 
\includegraphics[width=0.5\columnwidth]{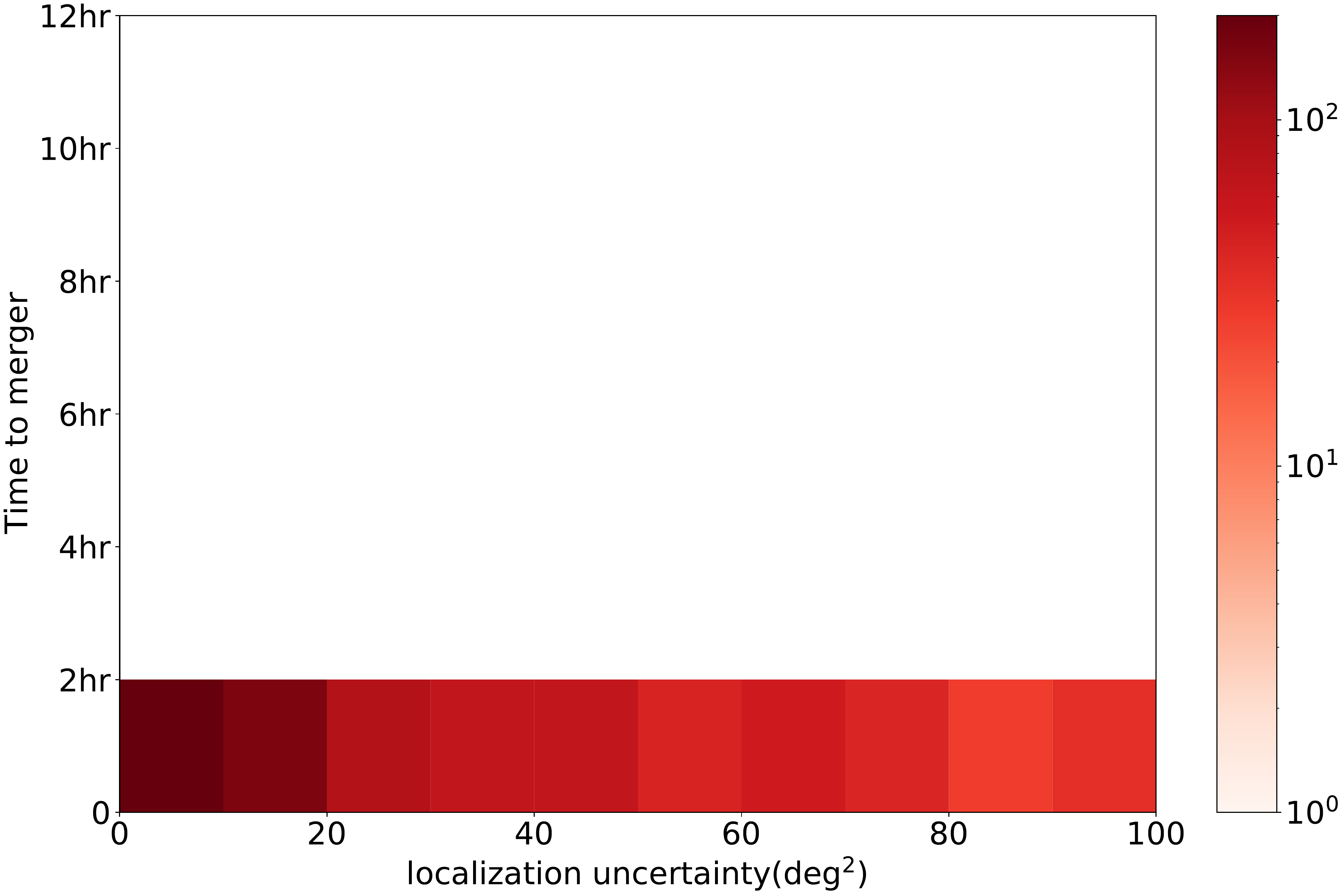}} 
\caption{Two-dimensional histograms showing the distributions of the time to merger
for \ac{BNS} mergers following \ac{DTD}. The horizontal axis of every subplot is the size of the $90\%$ credible region,  and the vertical axis represents the time to merger. 	
The color represents the number of sources which achieve 
the early warning criteria with the localisation requirement being $\leq 100\mathrm{deg^2}$ .}
\label{EW_3G_2d_100}
\end{figure}

    The cumulative distribution of localization uncertainty for \ac{BNS} mergers at 40 Mpc, 400 Mpc, 800 Mpc are shown in Figure~\ref{er_appendix}.   In addition, Figure~\ref{40EW}, Figure~\ref{400EW}, Figure~\ref{800EW}, Figure~\ref{1600EW} are the histogram distribution of time to merger respectively for \ac{BNS} mergers at 40 Mpc, 400 Mpc, 800 Mpc, 1600 Mpc. Finally, Figure~\ref{EW_3G_2d_100} show the two-dimensional histograms distributions of the time to merger
for \ac{BNS} mergers following \ac{DTD} with maximum allowable region of 100 deg$^2$.
      

  \end{appendices}


\begin{thebibliography}{}
\bibitem[Abadie et al.(2012)]{2012PhRvD..86f9903A} Abadie, J., Abbott, B.~P., Abbott, R., et al.\ 2012, \prd, 86, 069903
\bibitem[Abbott et al.(2016)]{2016PRL...116...061102L}B. P. Abbott et al. \ 2016, \prl, 116, 061102
\bibitem[Abbott et al.(2017)]{2017ApJ...850L..40A} Abbott, B.~P., Abbott, R., Abbott, T.~D., et al.\ 2017, \apjl, 850, L40. doi:10.3847/2041-8213/aa93fc
\bibitem[Abbott et al.(2017)]{2017CQGra..34d4001A} Abbott, B.~P., Abbott, R., Abbott, T.~D., et al.\ 2017, Classical and Quantum Gravity, 34, 044001. doi:10.1088/1361-6382/aa51f4
\bibitem[Abbott et al.(2017)]{2017PhRvL.119p1101A} Abbott, B.~P., Abbott, R., Abbott, T.~D., et al.\ 2017, \prl, 119, 161101
\bibitem[Abbott et al.(2019)]{2019PhRvX...9c1040A} Abbott, B.~P., Abbott, R., Abbott, T.~D., et al.\ 2019, Physical Review X, 9, 031040
\bibitem[Abbott et al.(2020)]{2020arXiv201014527A} Abbott, R., Abbott, T.~D., Abraham, S., et al.\ 2020, arXiv:2010.14527
\bibitem[Abbott et al.(2021)]{2021ApJ...909..218A} Abbott, B.~P., Abbott, R., Abbott, T.~D., et al.\ 2021, \apj, 909, 218. doi:10.3847/1538-4357/abdcb7
\bibitem[Abbott et al.(2021)]{2021ApJ...913L...7A} Abbott, R., Abbott, T.~D., Abraham, S., et al.\ 2021, \apjl, 913, L7. doi:10.3847/2041-8213/abe949
\bibitem[Acernese et al.(2015)]{2015CQGra..32b4001A} Acernese, F., Agathos, M., Agatsuma, K., et al.\ 2015, Classical and Quantum Gravity, 32, 024001. doi:10.1088/0264-9381/32/2/024001
\bibitem[Akcay(2018)]{2018arXiv180810057A} Akcay, S.\ 2018, Annalen der Physik, 531, 1. doi:10.1002/andp.201800365
\bibitem[Akcay et al.(2018)]{2018arXiv181207307A} Akcay, S., Fraser, M., \& Martin-Carrillo, A.\ 2018, arXiv:1812.07307
\bibitem[Akutsu et al.(2018)]{2018arXiv181108079A} Akutsu, T., Ando, M., Arai, K., et al.\ 2018, arXiv e-prints, arXiv:1811.08079
\bibitem[Berry et al.(2015)]{2015ApJ...804..114B} Berry, C.~P.~L., Mandel, I., Middleton, H., et al.\ 2015, \apj, 804, 114. doi:10.1088/0004-637X/804/2/114
\bibitem[Essick et al.(2017)]{2017PhRvD..96h4004E} Essick, R., Vitale, S., \& Evans, M.\ 2017, \prd, 96, 084004. doi:10.1103/PhysRevD.96.084004
\bibitem[ET science team (2011)]{2011ET-0106C-10} Amaro-Seoane P et al. 2009, \href{http://www.et-gw.eu/index.php/relevant-et-documents}{Einstein Telescope Conceptual design study.} ET internal note: ET-0106C-10
\bibitem[Barriga et al.(2010)]{2010CQGra..27h4005B} Barriga, P., Blair, D.~G., Coward, D., et al.\ 2010, Classical and Quantum Gravity, 27, 084005. doi:10.1088/0264-9381/27/8/084005
\bibitem[Bassett et al.(2011)]{2011IJMPD..20.2559B} Bassett, B.~A., Fantaye, Y., Hlozek, R., et al.\ 2011, International Journal of Modern Physics D, 20, 2559
\bibitem[Branchesi(2016)]{2016JPhCS.718b2004B} Branchesi, M.\ 2016, Journal of Physics Conference Series, 718, 022004. doi:10.1088/1742-6596/718/2/022004
\bibitem[Buonanno et al.(2009)]{2009PhRvD..80h4043B} Buonanno, A., Iyer, B.~R., Ochsner, E., et al.\ 2009, \prd, 80, 084043
\bibitem[Cannon et al.(2012)]{2012ApJ...748..136C} Cannon, K., Cariou, R., Chapman, A., et al.\ 2012, \apj, 748, 136. doi:10.1088/0004-637X/748/2/136
\bibitem[Chan et al.(2018)]{2018PhRvD..97l3014C} Chan, M.~L., Messenger, C., Heng, I.~S., et al.\ 2018, \prd, 97, 123014
\bibitem[Dwyer et al.(2015)]{2015PhRvD..91h2001D} Dwyer, S., Sigg, D., Ballmer, S.~W., et al.\ 2015, \prd, 91, 082001. doi:10.1103/PhysRevD.91.082001
\bibitem[Fairhurst(2009)]{2009NJPh...11l3006F} Fairhurst, S.\ 2009, New Journal of Physics, 11, 123006. doi:10.1088/1367-2630/11/12/123006
\bibitem[Fairhurst(2011)]{2011CQGra..28j5021F} Fairhurst, S.\ 2011, Classical and Quantum Gravity, 28, 105021. doi:10.1088/0264-9381/28/10/105021
\bibitem[Hachisu et al.(2008)]{2008ApJ...683L.127H} Hachisu, I., Kato, M., \& Nomoto, K.\ 2008, \apjl, 683, L127. doi:10.1086/591646
\bibitem[Hajela et al.(2019)]{2019ApJ...886L..17H} Hajela, A., Margutti, R., Alexander, K.~D., et al.\ 2019, \apjl, 886, L17. doi:10.3847/2041-8213/ab5226
\bibitem[Hotokezaka et al.(2019)]{2019NatAs...3..940H} Hotokezaka, K., Nakar, E., Gottlieb, O., et al.\ 2019, Nature Astronomy, 3, 940. doi:10.1038/s41550-019-0820-1
\bibitem[Howell et al.(2018)]{2018MNRAS.474.4385H} Howell, E.~J., Chan, M.~L., Chu, Q., et al.\ 2018, \mnras, 474, 4385. doi:10.1093/mnras/stx3077
\bibitem[Madau \& Dickinson(2014)]{2014ARA&A..52..415M} Madau, P. \& Dickinson, M.\ 2014, \araa, 52, 415
\bibitem[Maggiore(2007)]{Maggiore}M. Maggiore, Gravitational Waves. Vol. 1: Theory and Experiments, (Oxford University Press, Oxford, England, 2007)
\bibitem[Margutti et al.(2018)]{2018arXiv181204051M} Margutti, R., Cowperthwaite, P., Doctor, Z., et al.\ 2018, arXiv e-prints, arXiv:1812.04051
\bibitem[Metzger et al.(2010)]{2010MNRAS.406.2650M} Metzger, B.~D., Mart{\'\i}nez-Pinedo, G., Darbha, S., et al.\ 2010, \mnras, 406, 2650. doi:10.1111/j.1365-2966.2010.16864.x
\bibitem[Metzger \& Berger(2012)]{2012ApJ...746...48M} Metzger, B.~D., \& Berger, E.\ 2012, \apj, 746, 48
\bibitem[Mills et al.(2018)]{2018PhRvD..97j4064M} Mills, C., Tiwari, V., \& Fairhurst, S.\ 2018, \prd, 97, 104064. doi:10.1103/PhysRevD.97.104064
\bibitem[Nissanke et al.(2011)]{2011ApJ...739...99N} Nissanke, S., Sievers, J., Dalal, N., et al.\ 2011, \apj, 739, 99. doi:10.1088/0004-637X/739/2/99
\bibitem[Nuttall \& Sutton(2010)]{2010PhRvD..82j2002N} Nuttall, L.~K. \& Sutton, P.~J.\ 2010, \prd, 82, 102002. doi:10.1103/PhysRevD.82.102002
\bibitem[Piran(1992)]{1992ApJ...389L..45P} Piran, T.\ 1992, \apjl, 389, L45. doi:10.1086/186345
\bibitem[Planck Collaboration et al.(2016)]{2016A&A...594A..13P} Planck Collaboration, Ade, P.~A.~R., Aghanim, N., et al.\ 2016, \aap, 594, A13
\bibitem[Punturo et al.(2010)]{2010CQGra..27s4002P} Punturo, M., Abernathy, M., Acernese, F., et al.\ 2010, Classical and Quantum Gravity, 27, 194002
\bibitem[Punturo et al.(2014)]{2014ASSL..404..333P} Punturo, M., L{\"u}ck, H., \& Beker, M.\ 2014, Advanced Interferometers and the Search for Gravitational Waves, 333
\bibitem[Regimbau et al.(2012)]{2012PhRvD..86l2001R} Regimbau, T., Dent, T., Del Pozzo, W., et al.\ 2012, \prd, 86, 122001
\bibitem[Reitze et al.(2019a)]{2019BAAS...51c.141R} Reitze, D., LIGO Laboratory: California Institute of Technology, LIGO Laboratory: Massachusetts Institute of Technology, et al.\ 2019, \baas, 51, 141
\bibitem[Reitze et al.(2019b)]{2019BAAS...51g..35R} Reitze, D., Adhikari, R.~X., Ballmer, S., et al.\ 2019, \baas
\bibitem[Safarzadeh \& Berger(2019a)]{2019ApJ...878L..12S} Safarzadeh, M. \& Berger, E.\ 2019, \apjl, 878, L12. doi:10.3847/2041-8213/ab24df
\bibitem[Safarzadeh et al.(2019b)]{2019ApJ...878L..13S} Safarzadeh, M., Berger, E., Ng, K.~K.~Y., et al.\ 2019, \apjl, 878, L13
\bibitem[Safarzadeh et al.(2019c)]{2019ApJ...878L..14S} Safarzadeh, M., Berger, E., Leja, J., et al.\ 2019, \apjl, 878, L14. doi:10.3847/2041-8213/ab24e3
\bibitem[Sathyaprakash et al.(2011)]{2011arXiv1108.1423S} Sathyaprakash, B., Abernathy, M., Acernese, F., et al.\ 2011, arXiv e-prints, arXiv:1108.1423
\bibitem[Schutz(2011)]{2011CQGra..28l5023S} Schutz, B.~F.\ 2011, Classical and Quantum Gravity, 28, 125023. doi:10.1088/0264-9381/28/12/125023
\bibitem[The LIGO Scientific Collaboration et al.(2021)]{2021arXiv210801045T} The LIGO Scientific Collaboration, the Virgo Collaboration, Abbott, R., et al.\ 2021, arXiv:2108.01045
\bibitem[Vallisneri(2008)]{2008PhRvD..77d2001V} Vallisneri, M.\ 2008, \prd, 77, 042001. doi:10.1103/PhysRevD.77.042001
\bibitem[Vitale \& Evans(2017)]{2017PhRvD..95f4052V} Vitale, S. \& Evans, M.\ 2017, \prd, 95, 064052. doi:10.1103/PhysRevD.95.064052
\bibitem[Wen \& Chen(2010)]{2010PhRvD..81h2001W} Wen, L. \& Chen, Y.\ 2010, \prd, 81, 082001. doi:10.1103/PhysRevD.81.082001
\bibitem[Zhao(2018)]{2018SSPMA..48g9805Z} Zhao, W.\ 2018, Scientia Sinica Physica, Mechanica \& Astronomica, 48, 079805
\bibitem[Zhao \& Wen(2018)]{2018PhRvD..97f4031Z} Zhao, W. \& Wen, L.\ 2018, \prd, 97, 064031. doi:10.1103/PhysRevD.97.064031
\bibitem[Zheng \& Ramirez-Ruiz(2007)]{2007ApJ...665.1220Z} Zheng, Z. \& Ramirez-Ruiz, E.\ 2007, \apj, 665, 1220. doi:10.1086/519544
\end{thebibliography}
\end{document}